\documentclass[a4paper,11pt]{article}
\usepackage{jheppub} % for details on the use of the package, please
\usepackage[T1]{fontenc} % if needed
\usepackage{lmodern}
\usepackage{mathrsfs}
\usepackage{multirow}
\usepackage{array}
\usepackage{slashed}
\usepackage{bm}

\newcommand{\dd}{\mathrm{d}}
\newcommand{\p}{\mathrm{p}}
\newcommand{\n}{\mathrm{n}}
\newcommand{\E}{\mathrm{E}}
\newcommand{\M}{\mathrm{M}}
\newcommand{\cm}{\mathrm{cm}}

\title{\boldmath Study of the electromagnetic form factors of the nucleons in the timelike region}

\author[a,b]{Qin-He Yang,}
\emailAdd{yqh@hnu.edu.cn}
\author[a,b]{Di Guo,}
\emailAdd{diguo@hnu.edu.cn}
\author[a,b]{Ming-Yan Li,}
\emailAdd{20000621@hnu.edu.cn}
\author[a,b,1]{Ling-Yun Dai \note{Corresponding author.},}
\emailAdd{dailingyun@hnu.edu.cn}
\author[c]{Johann Haidenbauer,}
\emailAdd{j.haidenbauer@fz-juelich.de}
\author[d,c,e]{and Ulf-G. Mei{\ss}ner}
\emailAdd{meissner@hiskp.uni-bonn.de}

\affiliation[a]{School of Physics and Electronics, Hunan University, Changsha 410082, China}
\affiliation[b]{Hunan Provincial Key Laboratory of High-Energy Scale Physics and Applications, Hunan University, Changsha 410082, China}
\affiliation[c]{Institute for Advanced Simulation (IAS-4), Forschungszentrum J\"ulich, D-52425 J\"ulich, Germany}
\affiliation[d]{Helmholtz Institut f\"ur Strahlen- und Kernphysik and Bethe Center for Theoretical Physics, Universit\"at Bonn, D-53115 Bonn, Germany}
\affiliation[e]{Tbilisi State University, Tbilisi 0186, Georgia}

\abstract{The electromagnetic form factors $G_{\rm E}$ and $G_{\rm M}$ of the proton and neutron 
in the timelike region are extracted in a study of the processes $e^+ e^-\to \bar{p}p$ 
and $e^+ e^-\to \bar{n}n$. 
The reaction amplitude is evaluated within the distorted wave Born approximation, 
with the interaction of the antinucleon-nucleon ($\bar{N}N$) pair taken into account.
The latter is constructed within $SU(3)$ chiral effective field theory up to 
next-to-leading order.
An excellent description of the $e^+ e^-\to \bar{N}N$ data in the energy region from the $\bar{N}N$ threshold up to center-of-mass energies $E_{\rm cm}=2.2$~GeV is achieved. Results for the electromagnetic form factors 
$G_{\rm E}, G_{\rm M}$, $G_{\rm E}/G_{\rm M}$, and the 
subtracted effective form factors, $G_{\rm osc}$, are provided.
These can be helpful for further studies of the properties of the nucleons.}

\begin{document}
\allowdisplaybreaks[4]
\maketitle
\flushbottom

\section{Introduction}
\label{sec:intro}
%%%
The nucleons and their properties are of fundamental interest in nuclear and particle physics 
as they are the basic components of nuclei, which constitute common matter. 
The electromagnetic form factors (EMFFs) of the proton and the neutron play an 
important role in understanding the properties of the internal structure of nucleons 
as well as strong interactions, see, e.g., Refs.~\cite{Denig:2012by,Xia:2021agf} for 
reviews on recent progress. Indeed, the amplitudes of $eN\to eN$,  $e^+e^-\to \bar{N}N$ 
and $\bar{N}N\to e^+e^-$ can be written in terms of the electric and magnetic form 
factors: $G_\E$ and $G_\M$, i.e., the so-called EMFFs. 
The EMFFs in the timelike region have recently received considerable attention from 
the physics community \cite{Lin:2021xrc,Lin:2021umz,Qian:2022whn,Tomasi-Gustafsson:2022tpu,Yang:2022qoy,Chen:2023oqs,Cao:2021asd}. They can be studied through the annihilation of positron-electron 
into antinucleon-nucleon pairs and/or the inverse processes. 

%%%%
In the early stages, in the 1970s, the reaction
$e^+ e^- \to \bar{p}p$ was measured at center-of-mass energies $E_{\cm}=2.1$~GeV and the EMFFs of the proton were 
extracted in an experiment at ADONE \cite{Castellano:1973wh}. Subsequently, 
several measurements of the proton EMFFs were performed by other experimental groups, 
see, e.g., Refs.~\cite{Delcourt:1979ed,Bisello:1983at,DM2:1990tut,Antonelli:1993vz,Antonelli:1994kq,Antonelli:1998fv}.  
The study of proton EMFFs also benefitted from measurements on $\bar{p}p$ annihilating into $e^+ e^-$
pairs \cite{Bassompierre:1977ks}. 
On the other hand, measurements of neutron EMFFs are rather sparse. The Fenice collaboration 
studied the EMFFs in the process $e^+ e^- \to \bar{n}n$ \cite{Antonelli:1993vz,Antonelli:1998fv}.
%%%
Although those measurements launched a new era of this interesting field, they had limited statistics and a limited energy range. Fortunately, the situation has improved significantly in the 21st century. 
The BaBar, CMD-3, BESIII, and SND collaborations have performed various measurements of the EMFFs, 
and the uncertainties are now much smaller. See e.g., 
Refs.~\cite{BaBar:2005pon,BaBar:2013ves,CMD-3:2015fvi,CMD-3:2018kql,BES:2005lpy,BESIII:2015axk,BESIII:2019tgo,BESIII:2019hdp,BESIII:2021rqk} for the proton EMFFs, and Refs.~\cite{Achasov:2014ncd,Druzhinin:2019gpo,BESIII:2021tbq,SND:2022wdb,BESIII:2022rrg} for those of the neutron. 
Indeed, an intriguing phenomenon has been observed in the latest measurements of $e^+ e^-\to \bar{p}p$ \cite{BESIII:2021rqk} and $e^+ e^-\to \bar{n}n$ \cite{BESIII:2021tbq} by 
the BESIII collaboration. 
A phase difference in the oscillations of the EMFFs, $\Delta D=|D_p-D_n|=(125\pm 12)^\circ $ between 
the proton and the neutron has been found in the energy range of $E_{\cm} \in [2.00,3.08]$~GeV~\cite{BESIII:2021tbq}. 
This attracted further interest in timelike EMFFs, as it could be an important clue to 
reveal the internal structure and the interactions of nucleons.
A natural question following the experimental result is whether this phenomenon holds true close to the threshold as well. Specifically, since
the oscillation is observed in the effective EMFFs,  
does the phase difference still exists in the individual EMFFs of the nucleons? 
To answer these questions, one needs to study the EMFFs in the low energy region, that is, 
$E_{\cm} \in [2M_N,2.2~{\rm GeV}]$, where $M_N$ is the nucleon mass,  and extract the individual EMFFs. This is the primary goal  
of this paper. 

The situation around the $\bar{N}N$ threshold is still not clear, as it is not easy for 
experimentalists to perform measurements with sufficient statistics in this energy region.  
Nevertheless, the CMD-3 and SND collaborations published pertinent data on the cross sections 
for $e^+ e^-\to \bar{p}p$ and $e^+ e^-\to \bar{n}n$, respectively~\cite{CMD-3:2018kql,SND:2022wdb}. 
Their measurements, though still afflicted by large errors, are helpful for further analysis 
of the EMFFs.  
On the theoretical side, investigations in the near-threshold region require the 
inclusion of the $\bar{N}N$ interaction in the final state in order to be conclusive. 
Indeed, it has been shown in several studies that the $\bar{N}N$ final-state interaction 
(FSI) has non-negligible effects around the 
threshold\footnote{For discussions about FSI effects 
we refer to Refs.~\cite{Dai:2014zta,Yao:2020bxx,Wang:2023njt}.}. 
Since chiral effective field theory ($\chi$EFT) is presently the best tool to describe 
the interaction of baryons in the low-energy region, we will adopt this framework to 
generate an appropriate hadronic $\bar{N}N$ scattering amplitude. 
%%
%%%%%
Similar to what is done in our previous works and also by some other groups, we apply a 
two-step procedure to evaluate the reaction amplitude for $e^+e^-\to \bar{N}N$.  
%%%
The hadronic $\bar{N}N$ scattering amplitude is obtained by solving the Lippmann-Schwinger 
(LS) equation for an interaction potential derived within $SU(3)$ $\chi$EFT. 
Then, the amplitude of $e^+ e^-$ annihilating into $\bar{N}N$ pairs is computed 
based on the distorted wave Born approximation 
(DWBA)~\cite{Haidenbauer:2014kja,Dai:2017fwx,Haidenbauer:2020wyp,Yang:2022kpm}. 
This two-step procedure, combining $\chi$EFT and DWBA, as discussed above, has already been 
proven to be successful in studying the EMFFs, not only for nucleons but also for other 
baryons. See e.g., applications in the reactions 
$e^+e^- \leftrightarrow \bar{N}N$~\cite{Haidenbauer:2014kja,Yang:2022qoy}, 
$e^+e^-\to \bar{\Lambda}\Lambda$~\cite{Haidenbauer:2016won,Cao:2018kos}, 
$e^+e^-\to\Sigma^+\Sigma^-$,$\dots$~\cite{Haidenbauer:2020wyp,Yan:2023yff}, 
$e^+e^-\to \Lambda_c^+\Lambda_c^-$~\cite{Dai:2017fwx} and so on. 

%%%%%%%%%%

This paper is organized as follows: In Sec.~\ref{Sec:II}, we describe the calculation of the amplitude and cross section of the processes of $e^+ e^-\to \bar{p}p$ and $e^+ e^-\to \bar{n}n$, with the FSI 
in the $\bar{N}N$ system taken into account. Details of the derivation of the $\bar{N}N$ interaction potential from $SU(3)$ $\chi$EFT are given.
In Sec.~\ref{Sec:III}, we present our fit to the phase shifts of the coupled ${}^3S_1-{}^3D_1$
partial wave in the isospin basis and the experimental data sets for $\bar{N}N$
cross sections and differential cross sections, in which 
the low-energy constants (LECs) of $\chi$EFT as well as other parameters are fixed. 
Then, the individual EMFFs of the proton and the neutron from the threshold up to $2.2$~GeV 
are extracted, and the underlying physical interpretation is discussed.  
Finally, a brief summary is given in Sec.~\ref{Sec:IV}. Various technicalities are relegated to the appendices.

%%%%%%%%%%

\section{Theoretical framework}
\label{Sec:II}
\subsection{Formulas for the amplitude of $e^+e^-\to \bar{N}N$ and the EMFFs}
\begin{figure}[ht]
    \centering
    \includegraphics[width=0.5\linewidth]{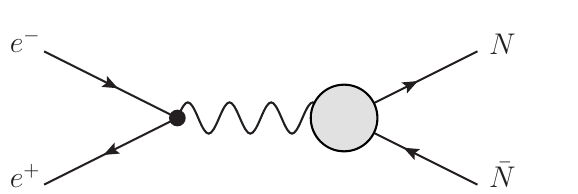}
    \caption{Illustrative diagram of the process $e^+ e^-\to \bar{N} N$.}
    \label{Fig:eeNN}
\end{figure}

%%%%%%%%% 
The differential cross section of the  processes $e^+ e^-\to \bar{p}p$, $\bar{n}n$,  is defined as \cite{Haidenbauer:2014kja}
\begin{align}
    \frac{\mathrm{d}\sigma}{\mathrm{d}\Omega}=\frac{1}{2s}\beta\, C(s)\sum_{i=1}^8|\phi_i|^2\,,  \label{eq:dsigma1}
\end{align}
where $\beta$ is the phase space factor, $\beta=k_N/k_e$ with $k_N$, $k_e$ the three-momenta of the
nucleon and electron in the center-of-mass frame (c.m.f.). The momenta are related to the total energy 
by $\sqrt{s}=2\sqrt{M_N^2+k_N^2}=2\sqrt{m_e^2+k_e^2}$, with $m_e$($M_N$) the electron (nucleon) mass. $C(s)$ is the S-wave Sommerfeld-Gamow factor, $C(y)=y/(1-e^{-y})$ 
with $y=\pi \alpha M_N/k_N$, where $\alpha=1/137.036$ is the fine-structure constant. 
For neutrons $C(y)\equiv 1$. 
The $\phi_i$'s are the standard helicity amplitudes for scattering of two spin-1/2 particles.
These are related to the ones for angular momentum helicity states by  
\begin{eqnarray}
    \langle \lambda'_1\lambda'_2|F|\lambda_1\lambda_2\rangle&=&\sum_{J}(2J+1)\langle \lambda'_1\lambda'_2|F^J|\lambda_1\lambda_2\rangle d^J_{\lambda\lambda'}(\theta) \,,\label{eq:helicity}
\end{eqnarray}
where $\lambda^{(\prime)}_{1,2}=\pm\frac{1}{2}$, $\lambda^{(\prime)}=\lambda^{(\prime)}_1-\lambda^{(\prime)}_2$, and the $d^J_{\lambda\lambda'}(\theta)$ are the Wigner $d$-functions.
To implement the information of the partial waves from $\chi$EFT, one needs to specify the 
quantum numbers. 
The transformation of the amplitudes from the helicity basis $|JM\lambda^{(\prime)}_1\lambda^{(\prime)}_2\rangle$ to the usual partial wave representation, $|JMLS\rangle$, is given by
\cite{Martin1970,Kuang:2023vac},
\begin{eqnarray}
    \langle \lambda'_1\lambda'_2|F^J|\lambda_1\lambda_2\rangle &=&\sum_{LS,L'S'}\frac{\sqrt{(2L+1)(2L'+1)}}{2J+1}\langle LS0\lambda|J\lambda\rangle \langle J\lambda'|L'S'0\lambda'\rangle \nonumber\\
    &&\quad \times \langle s_1s_2\lambda_1,-\lambda_2|S\lambda\rangle \langle S'\lambda'|s'_1s'_2\lambda'_1,-\lambda'_2\rangle \langle L'S'J|F|L SJ\rangle\,.\label{eq:JMLS}
\end{eqnarray}
%%%%%%%%%%%%%%%%%%%
%
In our analysis of $e^+e^-\to\bar{N}N$ the situation is simple because one has to deal only with
the $^3S_1- {}^3D_1$ partial wave\footnote{The amplitude written in Eq.~(\ref{eq:amp1}) is valid for 
one-photon exchange. 
Higher order terms of the electromagnetic vertex are neglected. 
Conservation of parity, charge conjugation, and time reversal invariance have been taken into account.}. 
The helicity amplitudes are decomposed into~\cite{Haidenbauer:2014kja,Buttimore:2006mq,Buttimore:2007cv}
\begin{eqnarray}
    \phi_1&=&\langle ++|F|++\rangle=\frac{\cos\theta}{2}[F_{00}-\sqrt{2}(F_{02}+F_{20})+2F_{22}]=-\frac{2m_eM_N\alpha}{s}\cos\theta G^N_{\E}\,,\nonumber\\
    \phi_2&=&\langle ++|F|--\rangle=\phi_1\,,\nonumber\\    
    \phi_3&=&\langle +-|F|+-\rangle=\frac{1+\cos\theta}{4}[F_{22}+2F_{00}+\sqrt{2}(F_{02}+F_{20})]=-\frac{\alpha}{2}(1+\cos\theta)G^N_{\M}\,,\nonumber\\
    \phi_4&=&\langle +-|F|-+\rangle=\frac{1-\cos\theta}{4}[F_{22}+2F_{00}+\sqrt{2}(F_{02}+F_{20})]=-\frac{\alpha}{2}(1-\cos\theta)G^N_{\M}\,,\nonumber\\
    \phi_5&=&\langle ++|F|+-\rangle=-\frac{\sin\theta}{2\sqrt{2}}[F_{02}-2F_{20}-\sqrt{2}(F_{22}-F_{00})]=\frac{M_N\alpha}{\sqrt{s}}\sin\theta G^N_{\E}\,,\nonumber\\
    \phi_6&=&\langle +-|F|++\rangle=\frac{\sin\theta}{2\sqrt{2}}[F_{20}-2F_{02}-\sqrt{2}(F_{22}-F_{00})]=-\frac{m_e\alpha}{\sqrt{s}}\sin\theta G^N_{\M}\,, \nonumber\\
    \phi_7&=&\langle ++|F|-+\rangle=-\phi_5\,,\nonumber\\   
    \phi_8&=&\langle -+|F|++\rangle=-\phi_6\,, \label{eq:amp1}
\end{eqnarray}
where $G^N_{\E}$ and $G^N_{\M}$ are the electric and magnetic nucleon form factors, respectively. $F_{L'L}$ is the partial wave amplitude of the process $e^+ e^-\to N\bar{N}$, 
as shown in Fig.~\ref{Fig:eeNN}.
It can be written as a product of two factors: one is the $e^+e^-\gamma$ vertex, and the other is the
$\bar{N}N\gamma$ effective vertex. 
Accordingly, the amplitude factorizes into 
\begin{align}
    F_{LL'}^{\bar{N}N,e^+e^-}=-\frac{4\alpha}{9}f_L^{\bar{N}N}f_{L'}^{e^+e^-} \nonumber\,,
\end{align}
with
\begin{eqnarray}
    &f_0^{\bar{N}N}=G_{\mathrm{M}}+\frac{M_N}{\sqrt{s}}G_{\mathrm{E}}\,,~~ &f_0^{e^+e^-}=1+\frac{m_e}{\sqrt{s}}\,, \nonumber\\
    &f_2^{\bar{N}N}=\frac{1}{\sqrt{2}}\left(G_{\mathrm{M}}-\frac{2M_N}{\sqrt{s}}G_{\mathrm{E}}\right)\,,~~ 
    &f_2^{e^+e^-}=\frac{1}{\sqrt{2}}\left(1-\frac{2m_e}{\sqrt{s}}\right)\,. \label{eq:vertex}
\end{eqnarray}
Obviously, one has $G_\E=G_\M$ at the $\bar{N}N$ thresholds which follows from
the definition of the Sachs form factors.
%%%%%%%%%%%%%%%%%%%%%%%%%%%%%%%
Inserting Eq.~(\ref{eq:amp1}) into Eq.~(\ref{eq:dsigma1}) leads to the
following formula for the differential cross 
section~\cite{BESIII:2021rqk,CMD-3:2018kql,Druzhinin:2019gpo}: 
\begin{eqnarray}\label{eq:dsigma2}
    \frac{\mathrm{d}\sigma}{\mathrm{d}\Omega}=\frac{\alpha^2\beta}{4s}C(s)\left[|G^N_{\mathrm{M}}(s)|^2(1+\cos^2\theta)+\frac{4M_N^2}{s}|G^N_{\mathrm{E}}(s)|^2\sin^2\theta\right] ,
\end{eqnarray}
where terms proportional to the electron mass $m_e$ have been ignored.
It reveals that the EMFFs can be extracted by analysing the dependence of the 
differential cross section on the scattering angle $\theta$. 
The reaction cross section is obtained by integrating Eq.~\eqref{eq:dsigma2} over the solid angle, 
\begin{eqnarray}
    \sigma=\frac{4\pi \alpha^2\beta}{3s}C(s)\left[|G^N_{\mathrm{M}}(s)|^2+\frac{2M_N^2}{s}|G^N_{\mathrm{E}}(s)|^2\right]\,. \label{eq:sigma}
\end{eqnarray}
%
%%%%%%%%%%%%%%%%%%%%%%%%%%%%%%%%%%%%

\subsection{Implementation of the $\bar NN$ FSI}
\begin{figure}[ht]
    \centering
    \includegraphics[width=0.9\linewidth]{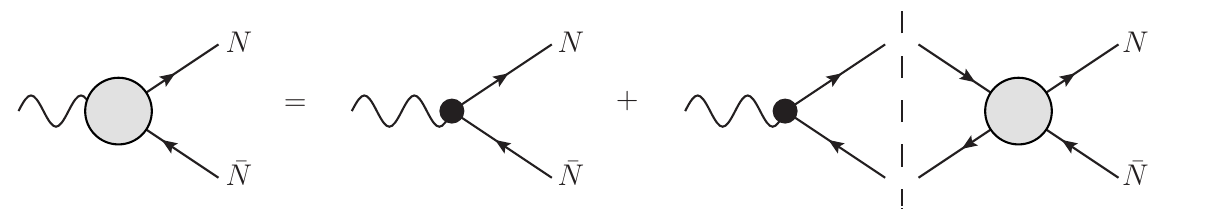}
    \caption{Diagrammatic representation of the vertex $\bar{N}N\gamma$  
 ($\gamma\to \bar{N}N$) consisting of the Born term
and re-scattering in the $\bar{N}N$ system.}
    \label{Fig:gNN}
\end{figure}

%%%%
As has been mentioned in the context of Eq.~(\ref{eq:vertex}), the vertex $\bar{N}N\gamma$
should include the FSI in the $\bar{N}N$ system. 
In the present work this is realized within the DWBA \cite{Haidenbauer:2014kja,Dai:2017fwx}:
\begin{eqnarray}
    f^{N\bar{N}}_{L}(k_N;E_{k_N}\!)=f^{N\bar{N},0}_{L}(k_N)\!+\!\sum_{L'}\!\int_0^{\infty}\!\frac{\dd pp^2}{(2\pi)^3}f^{N\bar{N},0}_{L'}(p)\frac{1}{2E_{k_N}-2E_p+i0^+}T_{L'L}(p,k_N;E_{k_N}\!)\,. \nonumber\\ \label{eq:DWBA}
\end{eqnarray}
A diagrammatic representation of Eq.~(\ref{eq:vertex}) can be found in Fig.~\ref{Fig:gNN},
where the full $\bar{N}N\gamma$ vertex is represented by the large shaded circle,
while the first diagram on the right-hand side is the so-called Born term, representing the 
bare $\bar{N}N\gamma$ production vertex $f^{\bar{N}N,0}_{L}$. 
We introduce two parameters to be fixed by fitting to the available experiments: 
$G^{\bar{p}p,0}_{\mathrm{E}}=G^{\bar{p}p,0}_{\mathrm{M}}$ and  $G^{\bar{n}n,0}_{\mathrm{E}}=G^{\bar{n}n,0}_{\mathrm{M}}$, with
\begin{eqnarray}
    &f_0^{\bar{N}N,0}(p)=G^{\bar{N}N,0}_{\mathrm{M}}+\frac{M_N}{2\sqrt{M_N^2+p^2}}G^{\bar{N}N,0}_{\mathrm{E}},~~ &f_2^{\bar{N}N,0}(p)=\frac{1}{\sqrt{2}}\left(G^{\bar{N}N,0}_{\mathrm{M}}-\frac{M_N}{\sqrt{M_N^2+p^2}}G^{\bar{N}N,0}_{\mathrm{E}}\right)\,. \label{eq:vtex;Born} \nonumber
\end{eqnarray}
It should be stressed that the parameters $G^{\bar{N}N,0}_{\mathrm{E}/\mathrm{M}}$ are complex due 
to possible intermediate annihilation processes, e.g., $\gamma\to\pi^+\pi^-\to \bar{N}N$. 
The second diagram in Fig.~\ref{Fig:gNN}, corresponding to the integral part of Eq.~(\ref{eq:DWBA}), 
provides the dressing of the vertex via $\bar{N}N$ re-scattering. 

The $\bar{N}N$ scattering amplitude $T_{LL'}(p,k_N;E_{k_N})$ in the relevant coupled ${}^3S_1-{}^3D_1$ 
partial wave will be calculated within the framework of $SU(3)$ $\chi$EFT. 
%%%
Indeed, there are many studies of the baryon-baryon and/or antibaryon-baryon interactions based on 
$\chi$EFT. For instance, Refs.~\cite{Epelbaum:2004fk,Epelbaum:2014efa,Epelbaum:1998ka,Epelbaum:1999dj} 
derived the $NN$ potential up to next-next-next-to-leading order (N$^3$LO) from $SU(2)$ $\chi$EFT. 
An extension to $SU(3)$ $\chi$EFT, describing the $NN$, hyperon-nucleon ($YN$), and 
hyperon-hyperon ($YY$) interactions, has been performed in 
Refs.~\cite{Haidenbauer:2013oca,Petschauer:2020urh,Haidenbauer:2009qn,Haidenbauer:2015zqb,Haidenbauer:2023qhf}, 
with potentials up to next-to-next-to-leading order (N$^2$LO) given in the literature. 
The merit of the present ansatz is that at low energies where FSI effects are expected to be 
essential the use of $\chi$EFT guarantees a reliable description of the $\bar NN$ interaction. 
With the extension to $SU(3)$ $\chi$EFT possible effects from hyperons, which appear 
as intermediate states, can be included. These could be of relevance because eventually 
we want to study the EMFFs up to 2.2~GeV which is close to the $\bar\Lambda\Lambda$
threshold. 
Nonetheless, it is clear that our calculation extends well beyond the usual and well-established
validity range of $\chi$EFT and that such an extension is mainly of phenomenological nature. 
However, it allows us to smoothly connect the region near the threshold, where the properties of
the EMFFs are strongly influenced by the $\bar NN$ FSI, with the region of $2-2.2$~GeV 
where a wealth of fairly precise data on the reactions $e^+e^-\to \bar pp$ and $e^+e^-\to \bar nn$ 
is available, including more selective observables like differential cross sections.

By exploiting the G-parity transformation the $\bar{N}N$ interaction 
can be easily derived from available $NN$ potentials, at least as far as the 
elastic part is concerned. 
For example, the $\bar{N}N$ potential up to N$^2$LO \cite{Kang:2013uia} utilizes the 
expressions for the $NN$ interaction of Ref.~\cite{Epelbaum:2004fk} as starting point 
and the N$^3$LO potential by Dai et al. \cite{Dai:2018tlc} is based on the $NN$ 
potential by Epelbaum et al.~\cite{Epelbaum:2014efa}. 
Indeed, for the interaction due to one boson exchange (OBE),
G-parity implies simply that   
\begin{eqnarray}\label{eq:GOBE}
    V^{\mathrm{OBE}}_{\bar{N}N}=(-1)^I V^{\mathrm{OBE}}_{NN},
\end{eqnarray}
where $I$ is the isospin of the exchanged pseudoscalar meson.
Also the contributions of two-boson exchange (TBE) involving pions and/or etas 
can be obtained through the G-parity transformation. However, it is not
applicable for contributions involving the $K$ and $\bar K$ mesons because these
do not have a well defined G-parity. In that case a separate calculation is necessary. 
For clarification, we work out the $\bar{N}N$ potential directly from $SU(3)$ $\chi$EFT,  
see the next subsection. 
Once the $\bar{N}N$ potential is established we insert it into the LS equation to obtain 
the scattering amplitude \cite{Kang:2013uia,Dai:2017ont}, 
\begin{align}
T_{L''L'}(p'',p';E_k)=V_{L''L'}(p'',p')+\sum_L\int\frac{\dd pp^2}{(2\pi)^3}V_{L''L}(p'',p)\frac{1}{2E_k-2E_p+i0^+}T_{LL'}(p,p';E_k), \label{eq:LS}
\end{align}
Here $E_k=\sqrt{s}/2$ and $V_{L''L}(p'',p')$ is the $\bar{N}N$ potential. 
Inserting $T_{L''L'}(p'',p';E_k)$ 
into Eq.~(\ref{eq:DWBA}), one can get the final amplitude for $e^+e^-\to \bar{N}N$. 
%
%%%

%%%%%%%%%%%%%%%%%%%%%%%%%%%%%%%%%%%%%%%%%%%%%%%%%%%%%

\subsection{The potential for the $\bar{N}N$ interaction}
\subsubsection{Lagrangians of $SU(3)$ $\chi$EFT}
The interaction Lagrangian of baryons coupling to mesons can be obtained from $SU(3)$ $\chi$EFT \cite{Bernard:1995dp,Haidenbauer:2013oca},
\begin{eqnarray}\label{eq:Lagrangian}
    \mathcal{L}_{MB}=\langle \bar{B}(i\gamma^\mu D_\mu-M_0)B\rangle-\frac{D}{2}\langle \bar{B}\gamma^\mu\gamma^5\{u_\mu,B\}\rangle-\frac{F}{2}\langle \bar{B}\gamma^\mu\gamma^5[u_\mu,B]\rangle\,,
\end{eqnarray}
where $\langle\cdots\rangle$ denotes the flavor trace, and the covariant derivative is 
$D_\mu B=\partial_\mu B+[\Gamma_\mu,B]$, with
$\Gamma_\mu=\frac{1}{2}(u^\dagger\partial_\mu u+u\partial_\mu u^\dagger)$. Here, one has 
$u_\mu=i(u^\dagger\partial_\mu u-u\partial_\mu u^\dagger)$ and  $u=\exp(i\Phi/\sqrt{2}f_0)$, 
where $f_0$ is the Goldstone boson decay constant in the three-flavor chiral limit, and $M_0$ is the baryon mass 
in the three-flavor chiral limit. $F$ and $D$ are coupling constants which satisfy 
$F+D=g_A\simeq1.27$. The baryon octet matrix $B$ and the meson octet matrix $\Phi$ have the 
following forms
\begin{eqnarray}
\Phi=\left(\begin{array}{ccc}
     \frac{\pi^0}{\sqrt{2}}+\frac{\eta}{\sqrt{6}}&\pi^+&K^+  \\
     \pi^-& -\frac{\pi^0}{\sqrt{2}}+\frac{\eta}{\sqrt{6}}&K^0\\
     K^-&\bar{K}^0&-\frac{2\eta}{\sqrt{6}}
\end{array}\right),\quad B=\left(\begin{array}{ccc}
     \frac{\Sigma^0}{\sqrt{2}}+\frac{\Lambda}{\sqrt{6}}&\Sigma^+&p  \\
     \Sigma^-& -\frac{\Sigma^0}{\sqrt{2}}+\frac{\Lambda}{\sqrt{6}}&n\\
     -\Xi^-&\Xi^0&-\frac{2\Lambda}{\sqrt{6}}
\end{array}\right)\,.
\end{eqnarray}
After expanding the Lagrangian in Eq.~(\ref{eq:Lagrangian}), the relevant interaction Lagrangians of the one- and two-mesons coupling with baryons can be obtained as 
\begin{eqnarray}\label{eq:BBphi}
    \mathcal{L}_{BB\Phi}&=&-\frac{\sqrt{2}}{2f_0}(D\langle \bar{B}\gamma^\mu\gamma^5\{\partial_{\mu}\Phi,B\}\rangle+F\langle \bar{B}\gamma^\mu\gamma^5[\partial_{\mu}\Phi,B]\rangle)\,,\nonumber\\
    \mathcal{L}_{BB\Phi\Phi}&=&\frac{i}{4f_0^2}\langle \bar{B}\left[[\Phi,\partial_\mu\Phi],B\right]\rangle\,.
\end{eqnarray}
The concrete form of the interaction Lagrangians can be obtained by taking the flavor trace, 
\begin{eqnarray}\label{eq:Lagrangian2}
    \mathcal{L}_{BB\Phi}&=&-f_{NN\pi}\left(\bar{p}\gamma^\mu\gamma^5p\partial_\mu\pi^0-\bar{n}\gamma^\mu\gamma^5n\partial_\mu\pi^0+\sqrt{2}\bar{p}\gamma^\mu\gamma^5n\partial_\mu\pi^++\sqrt{2}\bar{n}\gamma^\mu\gamma^5p\partial_\mu\pi^-\right)\nonumber\\
    &-&f_{\Lambda NK}\left(\bar{\Lambda}\gamma^\mu\gamma^5p\partial_\mu K^-+\bar{p}\gamma^\mu\gamma^5\Lambda\partial_\mu K^++\bar{\Lambda}\gamma^\mu\gamma^5n\partial_\mu \bar{K}^0+\bar{n}\gamma^\mu\gamma^5\Lambda\partial_\mu K^0\right)\nonumber\\
    &-&f_{\Sigma NK}\!\left(\!\bar{\Sigma}^0\gamma^\mu\gamma^5p\partial_\mu K^-\!+\!\bar{p}\gamma^\mu\gamma^5\Sigma^0\partial_\mu K^+ \!+\!\sqrt{2}\bar{\Sigma}^-\gamma^\mu\gamma^5p\partial_\mu \bar{K}^0 \!+\!\sqrt{2}\bar{p}\gamma^\mu\gamma^5\Sigma^+\partial_\mu K^0\right.\nonumber\\
    &&\left.~+\sqrt{2}\bar{\Sigma}^+\gamma^\mu\gamma^5n\partial_\mu K^-\!+\!\sqrt{2}\bar{n}\gamma^\mu\gamma^5\Sigma^-\partial_\mu K^+ \!-\!\bar{\Sigma}^0\gamma^\mu\gamma^5n\partial_\mu \bar{K}^0 \!-\!\bar{n}\gamma^\mu\gamma^5\Sigma^0\partial_\mu K^0 \!\right)\nonumber\\
    &-&f_{NN\eta}\left(\bar{p}\gamma^\mu\gamma^5p\partial_\mu\eta+\bar{n}\gamma^\mu\gamma^5n\partial_\mu\eta\right)+\cdots\,,\nonumber\\
    \mathcal{L}_{BB\Phi\Phi}&=&\frac{i}{4f_0^2}\left[\bar{p}\gamma^\mu p(\pi^+\partial_\mu\pi^--\pi^-\partial_\mu\pi^+)-\bar{n}\gamma^\mu n(\pi^+\partial_\mu\pi^--\pi^-\partial_\mu\pi^+)\right.\nonumber\\
    &-&\left.\sqrt{2}\bar{p}\gamma^\mu n(\pi^+\partial_\mu\pi^0-\pi^0\partial_\mu\pi^+)-\bar{n}\gamma^\mu p(\pi^0\partial_\mu\pi^--\pi^-\partial_\mu\pi^0)\right.\nonumber\\
    &+&\left.2\bar{p}\gamma^\mu p(K^+\partial_\mu K^--K^-\partial_\mu K^+)+\bar{n}\gamma^\mu n(K^+\partial_\mu K^--K^-\partial_\mu K^+)\right.\nonumber\\
    &+&\left.\bar{p}\gamma^\mu p(K^0\partial_\mu \bar{K}^0-\bar{K}^0\partial_\mu K^0)+2\bar{n}\gamma^\mu n(K^0\partial_\mu \bar{K}^0-\bar{K}^0\partial_\mu K^0)\right.\nonumber\\
    &+&\left.\bar{p}\gamma^\mu n(K^+\partial_\mu \bar{K}^0-\bar{K}^0\partial_\mu K^+)+\bar{n}\gamma^\mu p(K^0\partial_\mu K^--K^-\partial_\mu K^0)\right]+\cdots\,,
\end{eqnarray}
where \lq $\cdots$' represents terms that will not be used in the calculation of the $\bar{N}N$ scattering potentials up to NLO. The coupling constants are given as \cite{Haidenbauer:2013oca},
\begin{eqnarray}
    f_{NN\pi}&=&f,\quad f_{NN\eta}=\frac{1}{\sqrt{3}}(4\alpha-1)f,\nonumber\\
    f_{\Lambda NK}&=&-\frac{1}{\sqrt{3}}(1+2\alpha)f,\quad 
    f_{\Sigma NK}=(1-2\alpha)f\,,
\end{eqnarray}
with $\alpha=F/(F+D)$ and $f=g_A/2f_0$. The calculation will be performed in the framework of old-fashioned time-ordered perturbation theory \cite{Machleidt1987}, where the interaction Hamiltonians are needed and are defined as 
\begin{eqnarray}
    W_I=-\int\dd\bm{x}^3[\mathcal{L}_{I}(x)]_{x^0=0}\,,
\end{eqnarray}
where $\mathcal{L}_I$ is the Lagrangians of $\mathcal{L}_{BB\Phi}$ and $\mathcal{L}_{BB\Phi\Phi}$ as 
given in Eq.~\eqref{eq:Lagrangian2}. 

%%%%%%%%%%%%%%%%%%%%%%%%%%%%%%%%%%%%%%%%%%%%%%%%%%%

\subsubsection{The OBE potential}
\begin{figure}[htp]
    \centering
    \includegraphics[width=0.99\linewidth]{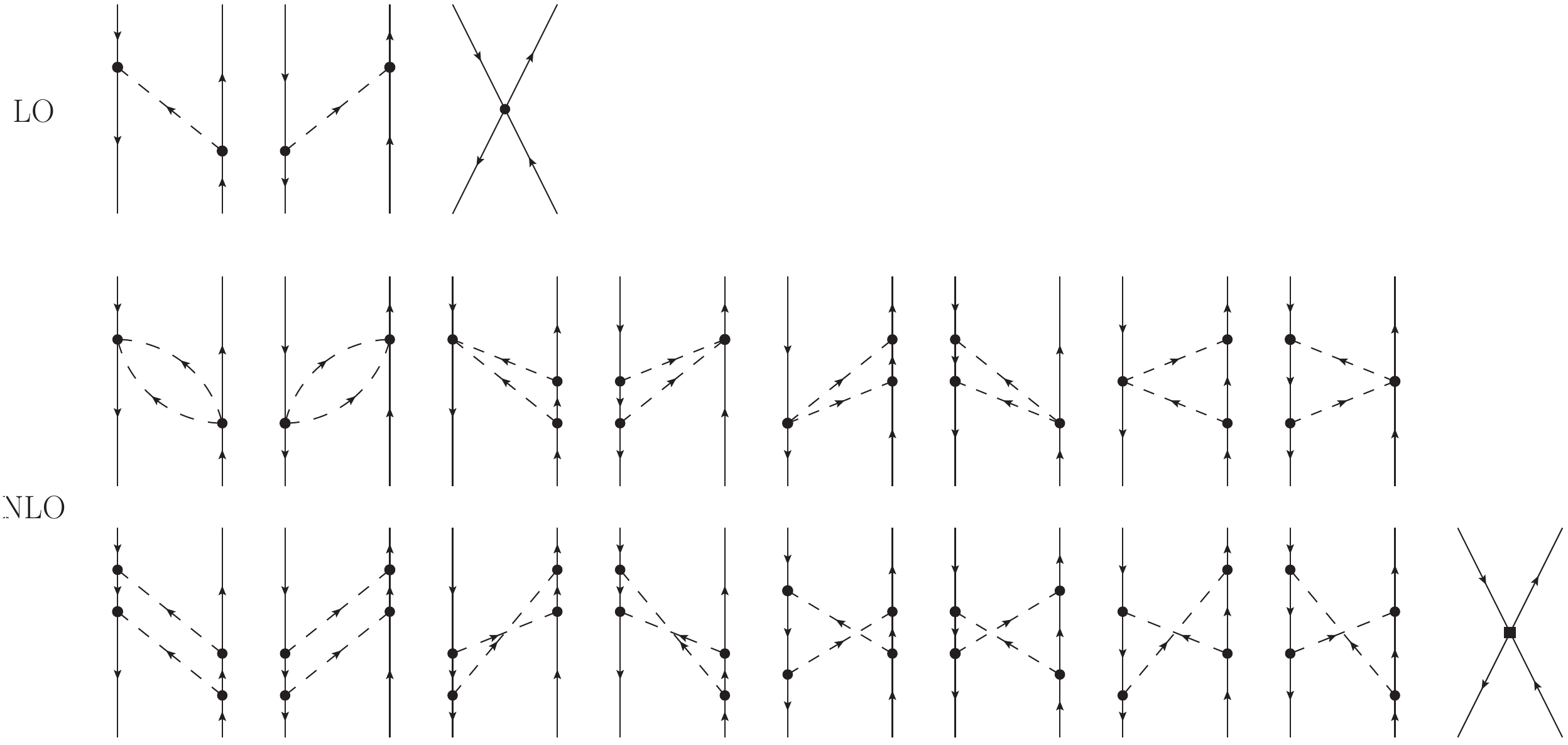}
    \caption{The Feynman diagram for $\bar{N}N$ scattering up to NLO.}
    \label{fig:NNbarFeynman}
\end{figure}

%%%%%%
In practice, the $\bar{N}N$ scattering equation is solved in the isospin basis. Hence, we first calculate the potentials in the physical basis according to time-ordered perturbation theory and 
then transform them into the isospin basis. 
In the physical basis, one has $V_{\bar{p}p\to\bar{p}p}=V_{\bar{n}n\to\bar{n}n}$ and $V_{\bar{p}p\to\bar{n}n}=V_{\bar{n}n\to\bar{p}p}$, neglecting  isospin breaking. 
The relation of the potentials between the physical and isospin bases satisfies 
\begin{eqnarray}\label{eq:PI}
    V_{\bar{p}p\to\bar{p}p}=\frac{1}{2}(V^{I=0}_{\bar{N}N}+V^{I=1}_{\bar{N}N}),\quad V_{\bar{p}p\to\bar{n}n}=\frac{1}{2}(V^{I=0}_{\bar{N}N}-V^{I=1}_{\bar{N}N})\,.
\end{eqnarray}
As a result, only $V_{\bar{p}p\to\bar{p}p}$ and $V_{\bar{p}p\to\bar{n}n}$ are needed to determine 
the whole potential. The contributions to the $\bar{N}N$ potential up to NLO are shown in 
Fig.~\ref{fig:NNbarFeynman}, 
and consist of OBEs, TBEs, and of contact terms. 
The TBE potentials include football, left and right triangle, planar, and cross box diagrams. 

The OBE potential for the process of $\bar{p}p\to\bar{p}p$ is defined as 
\begin{eqnarray}\label{eq:OBEamp}
\langle \bar{p}p|W_{BB\Phi}GW_{BB\Phi}|\bar{p}p\rangle=\frac{1}{(2\pi)^3}\delta^3(\bm{p}_1+\bm{p}_2-\bm{p}'_1-\bm{p}'_2)V^{\mathrm{OBE}}_{\bar{p}p\to\bar{p}p}\,,
\end{eqnarray}
where the propagator is $G=1/(z-H_0+i0^+)$, with $H_0$ the energy operator of the free particle and $z$ the total energy of the initial state. $\bm{p}_{1,2}$ is the three-momentum of the proton, antiproton in the initial state, and $\bm{p}'_{1,2}$ for the final state. In the language of creation and annihilation operators, the initial and final states are defined as 
\begin{eqnarray}\label{eq:IFstate}
|\bar{p}p\rangle=d^\dagger_{\bm{p}_1,\lambda_1}c^\dagger_{\bm{p}_2,\lambda_2}|0\rangle,\quad \langle \bar{p}p|=\langle 0|c_{\bm{p}'_1,\lambda'_1}d_{\bm{p}'_2,\lambda'_2}\,,
\end{eqnarray}
where $\lambda^{(')}_i\,(i=1,2)$ is the helicity of the proton or antiproton in initial (final) states. The potential in Eq.~\eqref{eq:OBEamp} includes $\pi^0$ and $\eta$ exchanges. Taking $\pi^0$ exchange as an example, the Lagrangian $-f_{NN\pi}\bar{p}\gamma^\mu\gamma^5p\partial_\mu\pi^0$ in Eq.~\eqref{eq:Lagrangian2} needs to be considered. The proton, anti\-proton and $\pi^0$ fields are defined as
\begin{eqnarray}\label{eq:field}
    \bar{p}&=&\frac{1}{(2\pi)^{3/2}}\sum_{\xi}\int \dd^3 \bm{k}\left[\bar{u}(\bm{k},\xi)c^\dagger_{\bm{k}\xi}e^{i k\cdot x}+\bar{v}(\bm{k},\xi)d_{\bm{k}\xi}e^{-i k\cdot x}\right]\,,\nonumber\\
    p&=&\frac{1}{(2\pi)^{3/2}}\sum_{\xi}\int \dd^3 \bm{k}\left[u(\bm{k},\xi)c_{\bm{k}\xi}e^{-i k\cdot x}+v(\bm{k},\xi)d^\dagger_{\bm{k}\xi}e^{i k\cdot x}\right]\,,\nonumber\\
    \pi^0&=&\frac{1}{(2\pi)^{3/2}}\int \frac{\dd^3\bm{l}}{\sqrt{2\omega_{l,\pi}}} \left(a^0_{\bm{l}}e^{-i l\cdot x}+a^{0\dagger}_{\bm{l}}e^{i l\cdot x}\right)\,,
\end{eqnarray}
where the pion energy is $\omega_{l,\pi}=\sqrt{l^2+m_\pi^2}$, with $l=|\bm{l}|$. The creation and annihilation operators of the proton and antiproton, $c^\dagger_{\bm{k}\xi}\,,d^\dagger_{\bm{k}\xi}$ and $c_{\bm{k}\xi}\,,d_{\bm{k}\xi}$, satisfy anticommutation relations, and the pion operators, $a^{0\dagger}_{\bm{l}}$ and $a^0_{\bm{l}}$ of $\pi^0$, satisfy  commutation relations. The spinors of the proton and the antiproton are given as  
\begin{eqnarray}\label{eq:uv}
    u(\bm{k},\xi)=\sqrt{\frac{E_k+M_N}{2E_k}}\left(\begin{matrix}
             1\\
            \frac{\bm{\sigma}\cdot \bm{k}}{E_k+M_N}
        \end{matrix}
        \right)|\xi\rangle,\quad v(\bm{k},\xi)=i\gamma^2 u^*(\bm{p},\xi)\,,
\end{eqnarray}
with the energy $E_k=\sqrt{k^2+M_N^2}$ and  $k=|\bm{k}|$. 
Applying Eqs.~\eqref{eq:IFstate}, \eqref{eq:field} and the commutation and anticommutation relations between the creation and annihilation operators, the $\pi^0$ exchange potential $V^{\pi^0}_{\bar{p}p\to\bar{p}p}$ for $\bar{p}p$ scattering can be obtained. One has 
\begin{eqnarray}\label{eq:Vpbp1}
    V^{\pi^0}_{\bar{p}p\to\bar{p}p}&=&f_{NN\pi}^2\left[-\frac{\bar{u}(\bm{p}'_2,\lambda'_2)\slashed{l}\gamma^5u(\bm{p}_2,\lambda_2)\bar{v}(\bm{p}_1,\lambda_1)\slashed{l}\gamma^5v(\bm{p}'_1,\lambda'_1)}{2\omega_{|\bm{p}'_1-\bm{p}_1|,\pi}(z-E_{p_2}-E_{p'_1}-\omega_{|\bm{p}'_1-\bm{p}_1|,\pi})}\right.\nonumber\\
    &&\left.-\frac{\bar{v}(\bm{p}_1,\lambda_1)\slashed{l}'\gamma^5v(\bm{p}'_1,\lambda'_1)\bar{u}(\bm{p}'_2,\lambda'_2)\slashed{l}'\gamma^5u(\bm{p}_2,\lambda_2)}{2\omega_{|\bm{p}'_1-\bm{p}_1|,\pi}(z-E_{p_1}-E_{p'_2}-\omega_{|\bm{p}'_1-\bm{p}_1|,\pi})}\right]+\cdots\,,
\end{eqnarray}
where $\cdots$ denotes the $s$ channel contribution caused by $\bar{N}N$ annihilating into a pion and then creating a $\bar{N}N$ pair. This part is not written out as it will be absorbed into the 
annihilation potential eventually. 
The corresponding Feynman diagrams are shown in the first row of Fig.~\ref{fig:NNbarFeynman}, the first term corresponds to the second diagram, with the momentum of $\pi^0$ given by $l^{\mu}=(\omega_{|\bm{p}'_1-\bm{p}_1|,\pi},-\bm{p}'_1+\bm{p}_1)$, and the second term corresponds to the first diagram, with $l'^{\mu}=(\omega_{|\bm{p}'_1-\bm{p}_1|,\pi},\bm{p}'_1-\bm{p}_1)$. 
According to Eq.~\eqref{eq:uv}, the relation between the $u$ and $v$ spinors can be obtained through transposition and Dirac matrix operation
%%%
\begin{eqnarray}\label{eq:vtou}
    \bar{v}\gamma^\mu\gamma^5v=-\bar{u}\gamma^{\mu}\gamma^{5}u\,.
\end{eqnarray}
%%%%
Transforming the terms in Eq.~\eqref{eq:Vpbp1} into the c.m.f. and applying Eq.~\eqref{eq:vtou}, 
the potential $V^{\pi^0}_{\bar{p}p\to\bar{p}p}$ is finally given as
\begin{eqnarray}\label{eq:pi0Amp}
V^{\pi^0}_{\bar{p}p\to\bar{p}p}&=&f_{NN\pi}^2\left[\frac{\bar{u}(-\bm{p}',\lambda'_2)\slashed{l}\gamma^5u(-\bm{p},\lambda_2)\bar{u}(\bm{p}',\lambda'_1)\slashed{l}\gamma^5u(\bm{p},\lambda_1)}{2\omega_{q,\pi}(z-E_{p}-E_{p'}-\omega_{q,\pi})}\right.\nonumber\\
    &&\left.+\frac{\bar{u}(\bm{p}',\lambda'_1)\slashed{l}'\gamma^5u(\bm{p},\lambda_1)\bar{u}(-\bm{p}',\lambda'_2)\slashed{l}'\gamma^5u(-\bm{p},\lambda_2)}{2\omega_{q,\pi}(z-E_{p}-E_{p'}-\omega_{q,\pi})}\right]\,,
\end{eqnarray}
where $\bm{p}$ and $\bm{p}'$ are the three-momenta of the initial and final states in the c.m.f., respectively.  Note that in the c.m.f., the momenta of the $\pi^0$ becomes 
$l^{\mu}=(\omega_{q,\pi},-\bm{q})$ and  $l'^{\mu}=(\omega_{q,\pi},\bm{q})$ for these two 
terms, where the transferred momentum is $\bm{q}=\bm{p}'-\bm{p}$. 
In the non-relativistic approximation the terms involving the spinors of
Eq.~\eqref{eq:uv} reduce to 
\begin{eqnarray}\label{eq:NonRelaApp}
    \bar{u}(\bm{p}',\lambda')\slashed{l}\gamma^5u(\bm{p},\lambda)\approx  \langle\lambda'|\bm{\sigma}\cdot \bm{q}|\lambda\rangle,\quad \bar{u}(\bm{p}',\lambda')\slashed{l}'\gamma^5u(\bm{p},\lambda)\approx - \langle\lambda'|\bm{\sigma}\cdot \bm{q}|\lambda\rangle.
\end{eqnarray}
Adopting the static approximation for $z$ in the propagator yields $z\simeq E_{p}+E_{p'}$ \cite{Machleidt1987}. 
Combining it with Eq.~\eqref{eq:NonRelaApp}, one obtains the final form of the one-pion exchange potential, 
\begin{eqnarray}
    V^{\pi^0}_{\bar{p}p\to\bar{p}p}\approx-f_{NN\pi}^2\frac{\langle\lambda'_2|\bm{\sigma}\cdot \bm{q}|\lambda_2\rangle \langle\lambda'_1|\bm{\sigma}\cdot \bm{q}|\lambda_1\rangle}{\bm{q}^2+m_\pi^2}\,.
\end{eqnarray}
Similarly, the OBE potential for the $\bar{p}p$ scattering from $\eta$ exchange has the following form
\begin{eqnarray}
    V^{\eta}_{\bar{p}p\to\bar{p}p}\approx-f_{NN\eta}^2\frac{\langle\lambda'_2|\bm{\sigma}\cdot \bm{q}|\lambda_2\rangle \langle\lambda'_1|\bm{\sigma}\cdot \bm{q}|\lambda_1\rangle}{\bm{q}^2+m_\eta^2}\,.
\end{eqnarray}

As mentioned above, the potential for the process $\bar{p}p\to\bar{n}n$ is also needed to determine 
the $\bar{N}N$ potential in the isospin basis. One has 
\begin{eqnarray}
    \langle \bar{n}n|W_{BB\Phi}GW_{BB\Phi}|\bar{p}p\rangle=\frac{1}{(2\pi)^3}\delta^3(\bm{p}_1+\bm{p}_2-\bm{p}'_1-\bm{p}'_2)V^{\mathrm{OBE}}_{\bar{p}p\to\bar{n}n}\,.
\end{eqnarray}
Here, only $\pi$ exchange appears. The relevant Lagrangians are $-f_{NN\pi}\sqrt{2}\bar{p}\gamma^\mu\gamma^5n\partial_\mu\pi^+$ and $-f_{NN\pi}\sqrt{2}\bar{n}\gamma^\mu\gamma^5p\partial_\mu\pi^-$ from $\mathcal{L}_{BB\Phi}$. With a calculation similar to the one above one obtains
\begin{eqnarray}
        V^{\mathrm{OBE}}_{\bar{\p}\p\to \bar{\n}\n}&=&-2f_{NN\pi}^2\frac{\langle\lambda'_2|\bm{\sigma}\cdot \bm{q}|\lambda_2\rangle \langle\lambda'_1|\bm{\sigma}\cdot \bm{q}|\lambda_1\rangle}{\bm{q}^2+m_\pi^2}\,.
\end{eqnarray}
Finally, the potential in the isospin basis can be obtained through the relation given in Eq.~\eqref{eq:PI},
\begin{eqnarray}\label{eq:VIso}
    V_{\bar{N}N}^{I=0}
    &=&-3f_{NN\pi}^2\frac{\langle\lambda'_2|\bm{\sigma}\cdot \bm{q}|\lambda_2\rangle \langle\lambda'_1|\bm{\sigma}\cdot \bm{q}|\lambda_1\rangle}{\bm{q}^2+m_\pi^2}-f_{NN\eta}^2\frac{\langle\lambda'_2|\bm{\sigma}\cdot \bm{q}|\lambda_2\rangle \langle\lambda'_1|\bm{\sigma}\cdot \bm{q}|\lambda_1\rangle}{\bm{q}^2+m_\eta^2}\,,\nonumber\\
V_{\bar{N}N}^{I=1}&=&f_{NN\pi}^2\frac{\langle\lambda'_2|\bm{\sigma}\cdot \bm{q}|\lambda_2\rangle \langle\lambda'_1|\bm{\sigma}\cdot \bm{q}|\lambda_1\rangle}{\bm{q}^2+m_\pi^2}-f_{NN\eta}^2\frac{\langle\lambda'_2|\bm{\sigma}\cdot \bm{q}|\lambda_2\rangle \langle\lambda'_1|\bm{\sigma}\cdot \bm{q}|\lambda_1\rangle}{\bm{q}^2+m_\eta^2}\,.
\end{eqnarray}
Defining an isospin factor $\mathcal{I}_{\bar{N}N\to\bar{N}N}$, similar to 
Ref.~\cite{Haidenbauer:2013oca}, the final results of OBE can be expressed as 
\begin{eqnarray}
    V^{\mathrm{OBE}}_{\bar{N}N\to\bar{N}N}=-f_{NNP}^2\frac{\langle\lambda'_2|\bm{\sigma}\cdot \bm{q}|\lambda_2\rangle \langle\lambda'_1|\bm{\sigma}\cdot \bm{q}|\lambda_1\rangle}{\bm{q}^2+m_P^2}\mathcal{I}_{\bar{N}N\to\bar{N}N}\,,
\end{eqnarray}
with $P$ the relevant pseudoscalar meson. 
The isospin factors can be extracted from the potentials as given in Eq.~\eqref{eq:VIso}, 
and they are listed in Table~\ref{tab:OBEIso}. 
\begin{table}[htp]
    \centering
    \begin{tabular}{|c|cc|}
        \hline
        Isospin&$\pi$&$\eta$\\\hline    
        $I=0$&  3 &1\\\hline
        $I=1$&$-$1&1\\\hline
    \end{tabular}
 \caption{The isospin factor $\mathcal{I}_{\bar{N}N\to\bar{N}N}$ for the OBE potentials.}
 \label{tab:OBEIso}
\end{table}
Comparing them with the isospin factors of the OBE potential of $NN$, as shown in table~2 
of Ref.~\cite{Haidenbauer:2013oca}, one finds signs/sign differences as expected  
from the G-parity transformation, Eq.~\eqref{eq:GOBE}.

%%%%%%%%%%%%%%%%%%%%%%%%%%%%%%%%%%%

\subsubsection{The TBE potential}\label{sec:TBE}
%%%%%%%%
The TBE potentials correspond to the set of one-loop Feynman diagrams shown in the second 
and third rows in Fig.~\ref{fig:NNbarFeynman}. 
As discussed above, the potentials will be taken as kernel in the LS equation to obtain the $\bar{N}N$ scattering amplitudes. Therefore, to avoid double counting, 
we should consider only the irreducible part of those  diagrams. 
In this section, we take the football diagrams as an example to illustrate the calculation of TBE potentials. The corresponding  Feynman diagrams are the first two graphs in the second row of Fig.~\ref{fig:NNbarFeynman}. The potential for the process $\bar{p}p\to\bar{p}p$ is defined as 
\begin{eqnarray}
    \langle \bar{p}p|W_{BB\Phi\Phi}GW_{BB\Phi\Phi}|\bar{p}p\rangle=\frac{1}{(2\pi)^3}\delta^3(\bm{p}_1+\bm{p}_2-\bm{p}'_1-\bm{p}'_2)V^{\mathrm{Football}}_{\bar{p}p\to\bar{p}p}\,,
\end{eqnarray}
where the potential $V^{\mathrm{Football}}_{\bar{p}p\to\bar{p}p}$ contains $\pi^+\pi^-$, $K^+K^-$ and $K^0\bar{K}^0$ exchanges. 
For $\pi^+\pi^-$ exchange, the vertex is coming from the chiral effective Lagrangian $i\bar{p}\gamma^\mu p(\pi^+\partial_\mu\pi^--\pi^-\partial_\mu\pi^+)/4f_0^2$. The $\pi^+$ and $\pi^-$ meson fields are defined as 
\begin{eqnarray}
    \pi^+&=&\frac{1}{(2\pi)^{3/2}}\int \frac{\dd^3\bm{l}}{\sqrt{2\omega_{l,\pi}}} \left(a_{\bm{l}}e^{-i l\cdot x}+b^{\dagger}_{\bm{l}}e^{i l\cdot x}\right)\,,\nonumber\\
    \pi^-&=&\frac{1}{(2\pi)^{3/2}}\int \frac{\dd^3\bm{l}}{\sqrt{2\omega_{l,\pi}}} \left(b_{\bm{l}}e^{-i l\cdot x}+a^{\dagger}_{\bm{l}}e^{i l\cdot x}\right)\,,
\end{eqnarray}
where $a^{\dagger}_{\bm{l}}\,,a_{\bm{l}}^{}$ and $b^{\dagger}_{\bm{l}}\,,b_{\bm{l}}^{}$ are the creation and annihilation operators of $\pi^+$ and $\pi^-$, respectively. After performing commutation and anti-commutation operations between the creation and annihilation operators for mesons and baryons, eliminating the Dirac delta function through momentum integration, and summing the spins, one has
\begin{eqnarray}\label{eq:Vpbp2}
    V^{\pi^+\pi^-}_{\bar{p}p\to\bar{p}p}&=&-\frac{1}{16f_0^4}\int\frac{\dd^3 \bm{l}_1}{(2\pi)^3}\left[\frac{\bar{u}(\bm{p}'_2,\lambda'_2)(\slashed{l}_1-\slashed{l}_2) u(\bm{p}_2,\lambda_2)\bar{v}(\bm{p}_1,\lambda_1)(\slashed{l}_1-\slashed{l}_2) v(\bm{p}'_1,\lambda'_1)}{4\omega_{l_1,\pi}\omega_{|\bm{l}_1+\bm{p}'_1-\bm{p}_1|,\pi}(z-E_{p_2}-E_{p'_1}-\omega_{l_1,\pi}-\omega_{|\bm{l}_1+\bm{p}'_1-\bm{p}_1|,\pi})}\right.\nonumber\\
    &&\left.+\frac{\bar{v}(\bm{p}_1,\lambda_1)(\slashed{l}_1-\slashed{l}'_2)v(\bm{p}'_1,\lambda'_1)\bar{u}(\bm{p}'_2,\lambda'_2)(\slashed{l}_1-\slashed{l}'_2) u(\bm{p}_2,\lambda_2)}{4\omega_{l_1,\pi}\omega_{|\bm{l}_1-\bm{p}'_1+\bm{p}_1|,\pi}(z-E_{p_1}-E_{p'_2}-\omega_{l_1,\pi}-\omega_{|\bm{l}_1-\bm{p}'_1+\bm{p}_1|,\pi})}\right]+\cdots\,,
\end{eqnarray}
where $\cdots$ denotes the $s$ channel contributions again. 
Specifically, the first diagram in the second row of Fig.~\ref{fig:NNbarFeynman} corresponds to the second term of Eq.~\eqref{eq:Vpbp2}, while the second diagram corresponds to the first term. The four-momenta of pions in the first term are $l_1^\mu=(\omega_{l_1,\pi},\bm{l}_1)$ for the $\pi^+$ and $l_2^\mu=(\omega_{|\bm{l}_1+\bm{p}'_1-\bm{p}_1|,\pi},-\bm{l}_1-\bm{p}'_1+\bm{p}_1)$ for the $\pi^-$, and the momenta in the second term are $l_1$ for the $\pi^+$ and  $l'^\mu_2=(\omega_{|\bm{l}_1-\bm{p}'_1+\bm{p}_1|,\pi},-\bm{l}_1+\bm{p}'_1-\bm{p}_1)$ for the $\pi^-$. 
One useful relation for transforming the Lorentz vectors composed of spinors and Dirac matrix is 
\begin{eqnarray}\label{eq:vtou2}
    \bar{v}\gamma^\mu v=\bar{u}\gamma^\mu u\,.
\end{eqnarray}
With it, one can transform the potentials given in Eq.~\eqref{eq:Vpbp2} into the c.m.f., 
\begin{eqnarray}
    V^{\pi^+\pi^-}_{\bar{p}p\to\bar{p}p}&=&-\frac{1}{16f_0^4}\int\frac{\dd^3 \bm{l}_1}{(2\pi)^3}\left[\frac{\bar{u}(-\bm{p}',\lambda'_2)(\slashed{l}_1-\slashed{l}_2) u(-\bm{p},\lambda_2)\bar{u}(\bm{p}',\lambda'_1)(\slashed{l}_1-\slashed{l}_2) u(\bm{p},\lambda_1)}{4\omega_{l_1,\pi}\omega_{|\bm{l}_1+\bm{q}|,\pi}(z-E_{p_2}-E_{p'_1}-\omega_{l_1,\pi}-\omega_{|\bm{l}_1+\bm{q}|,\pi})}\right.\nonumber\\
    &&\left.+\frac{\bar{u}(\bm{p}',\lambda'_1)(\slashed{l}_1-\slashed{l}'_2) u(\bm{p},\lambda_1)\bar{u}(-\bm{p}',\lambda'_2)(\slashed{l}_1-\slashed{l}'_2) u(-\bm{p},\lambda_2)}{4\omega_{l_1,\pi}\omega_{|\bm{l}_1-\bm{q}|,\pi}(z-E_{p_1}-E_{p'_2}-\omega_{l_1,\pi}-\omega_{|\bm{l}_1-\bm{q}|,\pi})}\right]+\cdots\,,
\end{eqnarray}
where $l_2^\mu=(\omega_{|\bm{l}_1+\bm{q}|,\pi},-\bm{l}_1-\bm{q})$ and $l'^\mu_2=(\omega_{|\bm{l}_1-\bm{q}|,\pi},-\bm{l}_1+\bm{q})$. For the spinor part in non-relativistic approximation one obtains 
\begin{eqnarray}
    &&\bar{u}(\bm{p}',\lambda')(\slashed{l}_1-\slashed{l}_2)u(\bm{p},\lambda)=\langle \lambda'|\lambda\rangle (\omega_{l_1,\pi}-\omega_{|\bm{l}_1+\bm{q}|,\pi})\,,\nonumber\\
    &&\bar{u}(\bm{p}',\lambda')(\slashed{l}_1-\slashed{l}'_2)u(\bm{p},\lambda)=\langle \lambda'|\lambda\rangle (\omega_{l_1,\pi}-\omega_{|\bm{l}_1-\bm{q}|,\pi})\,.
\end{eqnarray}
Then, the potential can be written in the form 
\begin{eqnarray}
    V^{\pi^+\pi^-}_{\bar{p}p\to\bar{p}p}&\approx& \frac{1}{16f_0^4}\int\frac{\dd^3 \bm{l}_1}{(2\pi)^3}\left[\frac{\langle \lambda'_2|\lambda_2\rangle\langle \lambda'_1|\lambda_1\rangle(\omega_{l_1,\pi}-\omega_{|\bm{l}_1+\bm{q}|,\pi})^2}{4\omega_{l_1,\pi}\omega_{|\bm{l}_1+\bm{q}|,\pi}(\omega_{l_1,\pi}+\omega_{|\bm{l}_1+\bm{q}|,\pi})}\right.\nonumber\\
    &&\left.+\frac{\langle \lambda'_2|\lambda_2\rangle\langle \lambda'_1|\lambda_1\rangle(\omega_{l_1,\pi}-\omega_{|\bm{l}_1-\bm{q}|,\pi})^2}{4\omega_{l_1,\pi}\omega_{|\bm{l}_1-\bm{q}|,\pi}(\omega_{l_1,\pi}+\omega_{|\bm{l}_1-\bm{q}|,\pi})}\right]\nonumber\\
    &=&\frac{1}{32f_0^4}\int\frac{\dd^3 \bm{l}_1}{(2\pi)^3}\frac{\langle \lambda'_2|\lambda_2\rangle\langle \lambda'_1|\lambda_1\rangle(\omega_{l_1,\pi}-\omega_{|\bm{l}_1+\bm{q}|,\pi})^2}{\omega_{l_1,\pi}\omega_{|\bm{l}_1+\bm{q}|,\pi}(\omega_{l_1,\pi}+\omega_{|\bm{l}_1+\bm{q}|,\pi})}\nonumber\\
    &=&\frac{1}{32f_0^4}\int\frac{\dd^3 \bm{l}_1}{(2\pi)^3}\frac{\langle \lambda'_2|\lambda_2\rangle\langle \lambda'_1|\lambda_1\rangle(\omega_{|\bm{l}_1-\bm{q}/2|,\pi}-\omega_{|\bm{l}_1+\bm{q}/2|,\pi})^2}{\omega_{|\bm{l}_1-\bm{q}/2|,\pi}\omega_{|\bm{l}_1+\bm{q}/2|,\pi}(\omega_{|\bm{l}_1-\bm{q}/2|,\pi}+\omega_{|\bm{l}_1+\bm{q}/2|,\pi})}\nonumber\\
    &=&\frac{1}{32f_0^4}\int\frac{\dd^3 \bm{l}_1}{8(2\pi)^3}\frac{\langle \lambda'_2|\lambda_2\rangle\langle \lambda'_1|\lambda_1\rangle(\omega_{|\bm{l}_1/2-\bm{q}/2|,\pi}-\omega_{|\bm{l}_1/2+\bm{q}/2|,\pi})^2}{\omega_{|\bm{l}_1/2-\bm{q}/2|,\pi}\omega_{|\bm{l}_1/2+\bm{q}/2|,\pi}(\omega_{|\bm{l}_1/2-\bm{q}/2|,\pi}+\omega_{|\bm{l}_1/2+\bm{q}/2|,\pi})}\nonumber\\
    &=&\frac{1}{128f_0^4}\int\frac{\dd^3 \bm{l}_1}{(2\pi)^3}\frac{\langle \lambda'_2|\lambda_2\rangle\langle \lambda'_1|\lambda_1\rangle(\omega_{-,\pi}-\omega_{+,\pi})^2}{\omega_{-,\pi}\omega_{+,\pi}(\omega_{-,\pi}+\omega_{+,\pi})}\,,
\end{eqnarray}
where one has 
\begin{eqnarray}  \omega_{\pm,P}=\sqrt{(\bm{l}_1\pm\bm{q})^2+4m^2_P}=2\omega_{|\bm{l}_1/2\pm\bm{q}/2|,P}\,.
\end{eqnarray}
In the first, second, and third steps, one performs the reflection  $\bm{l}_1\to-\bm{l}_1$, translation $\bm{l}_1\to \bm{l}_1-\bm{q}/2$, and the scaling $\bm{l}_1\to\bm{l}_1/2$, respectively. 
%%%%

The football diagrams due to $K^+K^-$ and $K^0\bar{K}^0$ exchanges involve the Lagrangians $i\bar{p}\gamma^\mu p(K^+\partial_\mu K^--K^-\partial_\mu K^+)/2f_0^2$ and $i\bar{p}\gamma^\mu p(K^0\partial_\mu \bar{K}^0-\bar{K}^0\partial_\mu K^0)/4f_0^2$. Analogous to what has been done for the two pions exchange abovve, one has 
\begin{eqnarray}
    V^{K^+K^-}_{\bar{p}p\to\bar{p}p}&=&\frac{1}{32f_0^4}\int\frac{\dd^3 \bm{l}_1}{(2\pi)^3}\frac{\langle\lambda'_2|\lambda_2\rangle\langle\lambda'_1|\lambda_1\rangle(\omega_{-,K}-\omega_{+,K})^2}{\omega_{-,K}\omega_{+,K}(\omega_{-,K}+\omega_{+,K})}\,,\nonumber\\
    V^{K^0\bar{K}^0}_{\bar{p}p\to\bar{p}p}&=&\frac{1}{4}V^{K^+K^-}_{\bar{p}p\to\bar{p}p}\,.
\end{eqnarray}
At last, one obtains the complete potential of the process of $\bar{p}p\to\bar{p}p$ from the football diagrams,
\begin{eqnarray}
    V^{\mathrm{Football}}_{\bar{p}p\to\bar{p}p}&=&\frac{1}{128f_0^4}\int\frac{\dd^3 \bm{l}_1}{(2\pi)^3}\frac{\langle\lambda'_2|\lambda_2\rangle\langle\lambda'_1|\lambda_1\rangle(\omega_{-,\pi}-\omega_{+,\pi})^2}{\omega_{-,\pi}\omega_{+,\pi}(\omega_{-,\pi}+\omega_{+,\pi})}\nonumber\\
    &&+\frac{5}{128f_0^4}\int\frac{\dd^3 \bm{l}_1}{(2\pi)^3}\frac{\langle\lambda'_2|\lambda_2\rangle\langle\lambda'_1|\lambda_1\rangle(\omega_{-,K}-\omega_{+,K})^2}{\omega_{-,K}\omega_{+,K}(\omega_{-,K}+\omega_{+,K})}\,. \label{Eq:fb;pp2pp}
\end{eqnarray}
%%%%
Similarly, the complete potential of the process of $\bar{p}p\to\bar{n}n$, from the football diagrams with $\pi\pi$ and $K\bar{K}$ exchanges, are given by  
\begin{eqnarray}
    V^{\mathrm{Football}}_{\bar{p}p\to\bar{n}n}&=&\frac{1}{64f_0^4}\int\frac{\dd^3 \bm{l}_1}{(2\pi)^3}\frac{\langle\lambda'_2|\lambda_2\rangle\langle\lambda'_1|\lambda_1\rangle(\omega_{-,\pi}-\omega_{+,\pi})^2}{\omega_{-,\pi}\omega_{+,\pi}(\omega_{-,\pi}+\omega_{+,\pi})}\nonumber\\
    &&+\frac{1}{128f_0^4}\int\frac{\dd^3 \bm{l}_1}{(2\pi)^3}\frac{\langle\lambda'_2|\lambda_2\rangle\langle\lambda'_1|\lambda_1\rangle(\omega_{-,K}-\omega_{+,K})^2}{\omega_{-,K}\omega_{+,K}(\omega_{-,K}+\omega_{+,K})}\,. \label{Eq:fb;pp2nn}
\end{eqnarray}
With Eqs.~(\ref{Eq:fb;pp2pp},\ref{Eq:fb;pp2nn}), one gets the potentials of the football diagrams in the isospin basis,
\begin{eqnarray}\label{eq:football3}
    V_{\bar{N}N}^{I=0}&=&\frac{3}{128f_0^4}\int\frac{\dd^3 \bm{l}_1}{(2\pi)^3}\frac{\langle\lambda'_2|\lambda_2\rangle\langle\lambda'_1|\lambda_1\rangle(\omega_{-,\pi}-\omega_{+,\pi})^2}{\omega_{-,\pi}\omega_{+,\pi}(\omega_{-,\pi}+\omega_{+,\pi})}\nonumber\\
    &&+\frac{3}{64f_0^4}\int\frac{\dd^3 \bm{l}_1}{(2\pi)^3}\frac{\langle\lambda'_2|\lambda_2\rangle\langle\lambda'_1|\lambda_1\rangle(\omega_{-,K}-\omega_{+,K})^2}{\omega_{-,K}\omega_{+,K}(\omega_{-,K}+\omega_{+,K})}\,,\nonumber\\
 V_{\bar{N}N}^{I=1}&=&-\frac{1}{128f_0^4}\int\frac{\dd^3 \bm{l}_1}{(2\pi)^3}\frac{\langle\lambda'_2|\lambda_2\rangle\langle\lambda'_1|\lambda_1\rangle(\omega_{-,\pi}-\omega_{+,\pi})^2}{\omega_{-,\pi}\omega_{+,\pi}(\omega_{-,\pi}+\omega_{+,\pi})}\nonumber\\
    &&+\frac{1}{32f_0^4}\int\frac{\dd^3 \bm{l}_1}{(2\pi)^3}\frac{\langle\lambda'_2|\lambda_2\rangle\langle\lambda'_1|\lambda_1\rangle(\omega_{-,K}-\omega_{+,K})^2}{\omega_{-,K}\omega_{+,K}(\omega_{-,K}+\omega_{+,K})}\,.
\end{eqnarray}
Again, we can  define the isospin factor $\mathcal{I}^{\mathrm{Football}}_{\bar{N}N\to\bar{N}N}$ for simplicity. One has 
\begin{eqnarray}\label{eq:VFootball}
    V^{\mathrm{Football}}_{\bar{N}N}=-\frac{1}{1024f_0^4}\int\frac{\dd^3 \bm{l}'_1}{(2\pi)^3}\frac{\langle\lambda'_2|\lambda_2\rangle\langle\lambda'_1|\lambda_1\rangle(\omega_{+,P}-\omega_{-,P})^2}{\omega_{+,P}\omega_{-,P}(\omega_{+,P}+\omega_{-,P})}\mathcal{I}^{\mathrm{Football}}_{\bar{N}N\to\bar{N}N}\,.
\end{eqnarray}
The isospin factors are extracted from Eq.~\eqref{eq:football3} and listed in Table. \ref{tab:Isospinfactor}.
%%%%%
%%%%----------------------------------------------------
\begin{table}[htp]
    \centering
    \begin{tabular}{| c|c|c|c c c c|}
    \hline
         & \multirow{2}{*}{Isospin} & \rule[-0.3cm]{0cm}{0.7cm}\multirow{2}{*}{\rule[-0.2cm]{0cm}{0.3cm}Intermediate} &\multirow{2}{*}{$\pi\pi$} & \multirow{2}{*}{$\pi\eta$} & \multirow{2}{*}{$\eta\eta$} & \multirow{2}{*}{$K\bar{K}$}  \\ 
          &  & baryons  &  &  &  &    \\    \hline
       \multirow{2}{*}{Football}  & $I=0$ & & $-$24&  -- &   --  & $-$48  \\ \cline{2-7}
                                   & $I=1$ & &  8  &   --  &   --  & $-$32 \\ \hline
        \multirow{6}{*}{Left Triangle}&\multirow{3}{*}{$I=0$}&$N$&12& -- & -- & --\\
                                       &                      &$\Lambda$&--&--&--& 6\\
                                       &                      &$\Sigma$&--&--&-- & 6\\ \cline{2-7}
                                       &\multirow{3}{*}{$I=1$}&$N$&$-$4 &--&--&--\\
                                       &                      &$\Lambda$&--&--&--&2\\
                                       &                      &$\Sigma$&--&--&--&10\\ \hline
        \multirow{6}{*}{Right Triangle}&\multirow{3}{*}{$I=0$}&$\bar{N}$&12& -- & -- & --\\
                                       &                      &$\bar{\Lambda}$&--&--&--& 6\\
                                       &                      &$\bar{\Sigma}$&--&--&-- & 6\\ \cline{2-7}
                                       &\multirow{3}{*}{$I=1$}&$\bar{N}$&$-$4 &--&--&--\\
                                       &                      &$\bar{\Lambda}$&--&--&--&2\\
                                       &                      &$\bar{\Sigma}$&--&--&--&10\\ \hline 
        \multirow{8}{*}{Planar Box}    &\multirow{4}{*}{$I=0$}&$\bar{N}N$   &9& 6& 1&--\\
                                       &                      &$\bar{\Lambda}\Lambda$&--&--&--&2\\
                                       &                      &$\bar{\Sigma}\Sigma$&--&--&--& 6\\
                                       &                      & $\bar{\Sigma}\Lambda+\bar{\Lambda}\Sigma$&--&--&--&0\\ \cline{2-7}
                                       &\multirow{4}{*}{$I=1$}&$\bar{N}N$&1&$-$2&1&--\\
                                       &                      &$\bar{\Lambda}\Lambda$&--&--&--&0\\
                                       &                      &$\bar{\Sigma}\Sigma$&--&--&--&4\\
                                       &                      &$\bar{\Sigma}\Lambda+\bar{\Lambda}\Sigma$&--&--&--&4\\ \hline
        \multirow{8}{*}{Cross Box}    &\multirow{4}{*}{$I=0$}&$\bar{N}N$   &$-$3& 6& 1&--\\
                                       &                      &$\bar{\Lambda}\Lambda$&--&--&--&0\\
                                       &                      &$\bar{\Sigma}\Sigma$&--&--&--& 0\\
                                       &                      & $\bar{\Sigma}\Lambda+\bar{\Lambda}\Sigma$&--&--&--&0\\ \cline{2-7}
                                       &\multirow{4}{*}{$I=1$}&$\bar{N}N$&5&$-$2&1&--\\
                                       &                      &$\bar{\Lambda}\Lambda$&--&--&--&0\\
                                       &                      &$\bar{\Sigma}\Sigma$&--&--&--&0\\
                                       &                      &$\bar{\Sigma}\Lambda+\bar{\Lambda}\Sigma$&--&--&--&0\\ \hline                                               
    \end{tabular}
    \caption{Isospin factors of TBE potentials for $\bar{N}N$ scattering.}
    \label{tab:Isospinfactor}
\end{table}
%%%%----------------------------------------------------
Note that the isospin factors for the planar box, crossed box, left triangle, right triangle diagrams are also listed. 
A detailed derivation of those contributions can be found in Appendix \ref{app:TBE}. 

Comparing with the isospin factors of the $NN$ potentials given by Ref.~\cite{Haidenbauer:2013oca}, 
for $\pi$ and $\eta$ we recover again the G-parity transformation, that is, 
\begin{eqnarray}\label{eq:TBEG}
    V^{\mathrm{TBE}}_{\bar{N}N}=(-1)^{I_1+I_2}V^{\mathrm{TBE}}_{NN}\,, \nonumber
\end{eqnarray}
where $I_1$ and $I_2$ are the isospin of the two exchanged pseudoscalar bosons. 
Specifically, for $\pi\eta$ exchanges, the TBE potentials of $N\bar{N}$ and 
$NN$ scatterings differ by a minus sign, and for $\pi\pi$ and $\eta\eta$ exchanges, they have the 
same sign\footnote{It is worth pointing out that these relations have already been given in 
Ref.~\cite{Martin1970}, too.}. 
However, this rule is not applicable for $K\bar K$ exchange simply because $K$  and $\bar K$ 
do not have definite G-parity.

%%%%%%%%%%%%%%%%%%%%%%%%%%%%%%%%%%%%%%%%%%%%%%%%%%%%%%%

\subsection{The contact terms and the annihilation potential}
Besides the contributions from boson exchanges, there are standard contact terms and 
an annihilation part, too. 
The contact terms for $^3S_1-{}^3D_1$ partial waves up to NLO are given by~\cite{Dai:2017ont,Epelbaum:2014efa,Kang:2013uia}
\begin{eqnarray}\label{contact}
    V({}^3S_1)&=&\tilde{C}_{^3S_1}+C_{^3S_1}(p^2+p'^2)\,,\nonumber\\
    V({}^3D_1-{}^3S_1)&=&C_{\epsilon_1}p'^2,\quad V({}^3S_1-{}^3D_1)=C_{\epsilon_1}p^2\,,
\end{eqnarray}
where $p$ and $p'$ are the momenta in the c.m.f. of the initial and final $\bar{N}N$ systems, respectively. Here, the $\tilde{C}_i$ denote the LECs that arise at leading order (LO), corresponding to the contact terms without derivatives. The $C_i$ arise at NLO, corresponding to 
the terms with two derivatives. An essential difference between the $NN$ and $\bar{N}N$ interaction is the presence 
of annihilation processes in the latter. Following Refs.~\cite{Kang:2013uia,Dai:2017ont}, the annihilation part of the 
$\bar{N}N$ potential for the coupled $^3S_1-{}^3D_1$ partial wave is parameterized as
\begin{eqnarray}\label{ann}
    V^{\mathrm{ann}}({}^3S_1)&=&-i(\tilde{C}^a_{^3S_1}+C^a_{^3S_1}p^2)(\tilde{C}^a_{^3S_1}+C^a_{^3S_1}p'^2)\,,\nonumber\\
    V^{\mathrm{ann}}({}^3S_1-{}^3D_1)&=&-iC_{\epsilon_1}^ap'^2(\tilde{C}^a_{^3S_1}+C^a_{^3S_1}p^2)\,,\nonumber\\
    V^{\mathrm{ann}}({}^3D_1-{}^3S_1)&=&-iC_{\epsilon_1}^ap^2(\tilde{C}^a_{^3S_1}+C^a_{^3S_1}p'^2)\,,\nonumber\\
    V^{\mathrm{ann}}({}^3D_1)&=&-i[(C_{\epsilon_1}^a)^2+(C_{^3D_1}^a)^2]p^2p'^2 \,,
\end{eqnarray}
which is consistent with requirements from unitarity.
In the expressions above, the parameters $\tilde{C}^a$ and $C^a$ are real. All the potentials 
used in the LS equation are cut off by a regulator function, $f_R(\Lambda)=\exp[-(p^6+p'^6)/\Lambda^6]$, 
to suppress high-momentum components~\cite{Epelbaum:2004fk,Haidenbauer:2013oca}. To explore the
dependence of our results on the cutoff, we consider a range of cut-off values: $\Lambda=$750, 800, 850, 900, 950~MeV.  
These values are noticeably larger than the ones required and used in standard $\chi$EFT calculations 
\cite{Dai:2017ont,Epelbaum:2014efa}, but reflect the fact that we want to apply our $\bar NN$ interaction over an extended
energy range. 
%%%

%%%%%%%%%%%%%%%%%%%%%%%%%%%%%%%%%%%%%%%%%%%%%%%%%%%%%%%%%%%%%%%%%%%

\section{Results and discussion} 
\label{Sec:III} 

%%%%%%%%%%%%%%%%%%%%%%%%%%%%%%%
\subsection{Fit procedure}
The hadronic scattering amplitudes are taken as input in Eq.~(\ref{eq:DWBA}) to evaluate 
the electron-positron annihilation amplitude within the DWBA approach. 
Note that the amplitudes $T_{LL'}$ of $\bar{N}N$ scattering and $f^{\bar{N}N,I}_L$ of the $\bar{N}N\gamma$ vertex
are obtained in the isospin basis. To fit the experimental data sets, one needs to transform these amplitudes to the physical basis, 
\begin{equation}
   f^{\bar{p}p}_L=\frac{1}{\sqrt{2}}(f^{\bar{N}N,I=0}_L-f^{\bar{N}N,I=1}_L),\quad f^{\bar{n}n}_L=\frac{1}{\sqrt{2}}(f^{\bar{N}N,I=0}_L+f^{\bar{N}N,I=1}_L)\,.
\end{equation}
In the isospin basis, we use $M_N=(M_p+M_n)/2$, while in the calculation of the observables, such as differential cross section 
and cross sections, we take the physical masses in the phase-space factors.  
By fitting to the experimental data, the $e^+e^-\to \bar{N}N$ amplitudes are determined. With 
these amplitudes, the EMFFs $G_{\mathrm{M}}$ and $G_{\mathrm{E}}$ can be extracted through 
Eqs.~(\ref{eq:vertex},\ref{eq:dsigma2}). It is worth pointing out that the EMFFs 
of the nucleons very near to the threshold are true predictions.

To fix the LECs of $\chi$EFT and other unknown parameters such as $G^{N,0}_{\E,\M}$, we 
perform an overall fit to two kinds of data sets: One of them are the $S$-matrix elements from a
partial-wave analysis
(PWA) of $\bar{N}N$ scattering \cite{Zhou:2012ui} and $\bar NN$ scattering lengths from $\chi$EFT \cite{Kang:2013uia,Dai:2017ont}. 
The other set includes the cross 
sections, angular distributions, and EMFFs of the processes of $e^+e^-\to \bar{N}N$ as well as 
$\bar{N}N\to e^+e^-$. With regard to $\bar{N}N$ scattering we focus on energies not
too far from the threshold, because here we expect that $\chi$EFT works reliably. 
Specifically, we consider only the first three momenta of the PWA \cite{Zhou:2012ui} which means the
region $T_{\rm lab} \leq 50$~MeV. 
%%%   
There are 7 LECs in the $\bar{N}N$ interaction for each of the two isospin channels, 
see Eqs.~(\ref{contact},\ref{ann}). Since
they are all real numbers it implies a total of 14 LECs for the analysis.
There are two more parameters, related to the Born term of the EMFFs,  
namely $G_{\E/\M}^{0(I=0)}$ and $G_{\E/\M}^{0(I=1)}$. 
Keep in mind that the overall phase of the amplitude is not an observable. Thus, it 
can not be determined in fitting  the data. 
Therefore, one of the couplings of the EMFFs can be fixed to be real. 
Here, $G_{\E/\M}^{0(I=0)}$ is chosen to be real. In addition, there are 9 normalization 
factors to fit the number of events of the angular distributions, as the efficiencies are not known.  
Thus, there are 26 parameters to be determined in total. On the other hand, there are 722 
data points used in the analysis, including 154 cross-section 
values~\cite{BES:2005lpy,BESIII:2019hdp,BESIII:2019tgo,BESIII:2021rqk,BESIII:2021tbq,BaBar:2005pon,BaBar:2013ves,CMD-3:2015fvi,CMD-3:2018kql,Achasov:2014ncd,Druzhinin:2019gpo,Bardin:1994am,SND:2022wdb}, 
477 points of differential cross sections, 7 cross sections ratios 
($\sigma(e^+e^-\to \bar nn)/\sigma(e^+e^-\to \bar pp)$),
44 individual EMFFs ($G_\M$ and $G_\E$ that are extracted from the experiments), 
36 $S$-matrix elements from the $\bar NN$ PWA~\cite{Zhou:2012ui} and 2 $\bar NN$ scattering lengths \cite{Kang:2013uia,Dai:2017ont}. 
As will be discussed in the following subsections, the parameters 
can be fixed well due to the availability of a large data set.

%%%%%%%%%%%%%%%%%%%%%%%%%%%%%%%%%%%%%%%%%%%%%%%%%%%%%%%%%%%%%%%%%%%%%

\subsection{Fit to the $\bar{N}N$ scattering amplitudes}
\begin{figure}[pht]
    \centering
    \includegraphics[width=0.9\linewidth]{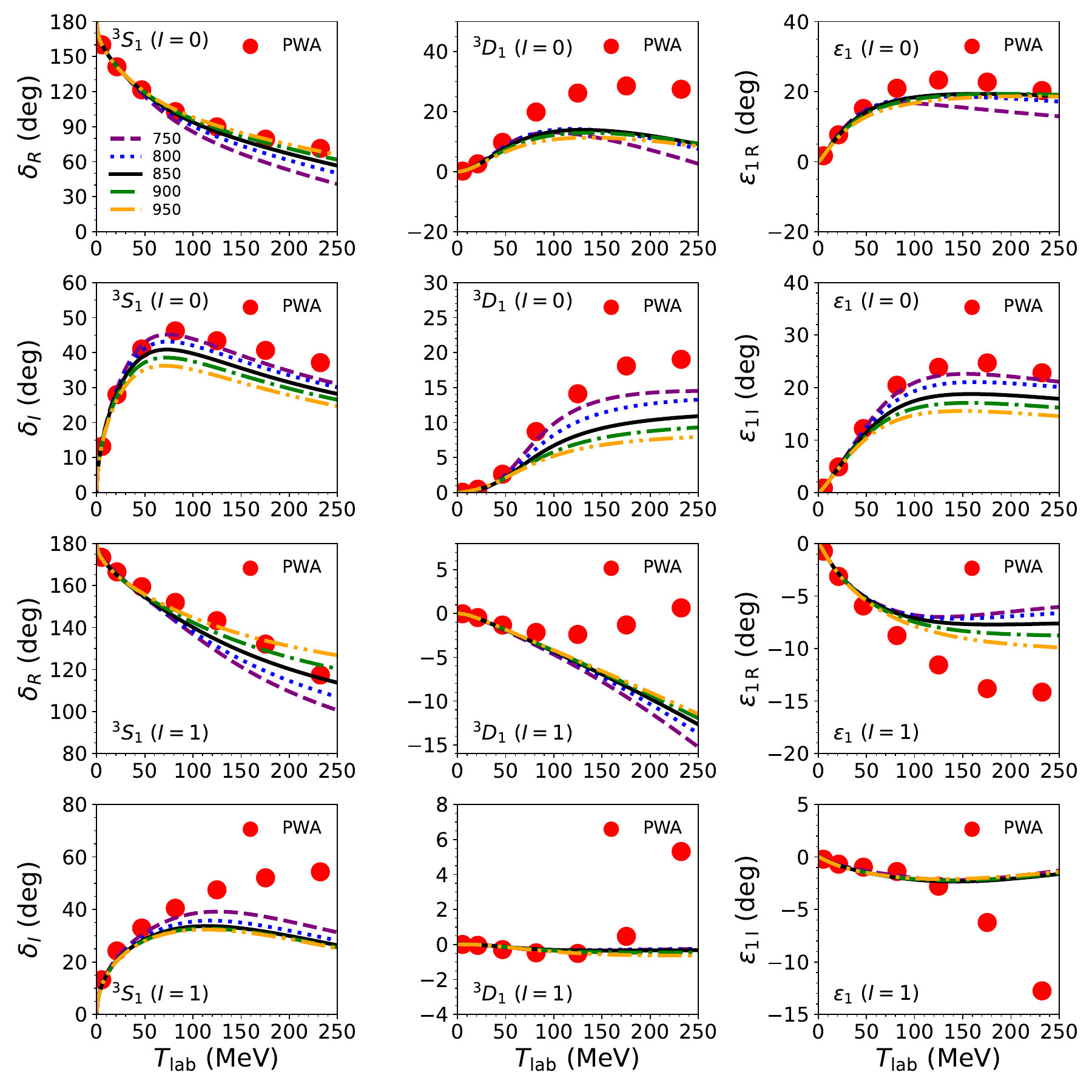}
    \caption{Real and imaginary parts of the phase shifts and inelasticities at NLO for the ${}^3S_1-{}^3D_1$ partial waves,  with isospin $I=0$ and/or $I=1$. The red-filled circles are the results of the PWA~\cite{Zhou:2012ui}. The green dashed, blue dotted, black solid, purple dash-dotted, and orange dash-dot-dotted lines represent our results with cutoffs $\Lambda=$750, 800, 850, 900, 950 MeV, respectively.   
    \label{Fig:PSdiffcutoff} }
\end{figure}

For simplicity, we will discuss the fit results of $\bar{N}N$ scattering in this subsection and 
the ones for $e^+e^-\to \bar NN$ in the next subsection. 
The $\bar{N}N$ potentials are calculated within $SU(3)$ $\chi$EFT up to NLO, in which there are 14 LECs 
corresponding to contact terms in the elastic and the annihilation parts. 
To fix these parameters we fit to the $S$-matrix elements obtained in a PWA 
of $\bar{N}N$ scattering data~\cite{Zhou:2012ui}. 
The phase shifts for the coupled $\bar{N}N$ partial wave $^3S_1-{}^3D_1$ can be 
extracted from the $S$-matrix as follows
\begin{eqnarray}
    \left(\begin{array}{cc}
        S_{00}&S_{02}\\
        S_{20}&S_{22}
    \end{array}\right)=\left(\begin{array}{cc}
    \cos(2\epsilon_1)e^{i2\delta_{0}}&-i\sin(2\epsilon_1)e^{i(\delta_{0}+\delta_{2})}\\
    -i\sin(2\epsilon_1)e^{i(\delta_{0}+\delta_{2})}&\cos(2\epsilon_1)e^{i2\delta_{2}}
\end{array}\right)\, ,  \label{eq:pscouple}
\end{eqnarray}
where we adapt the convention of Refs.~\cite{Dai:2017ont,Kang:2013uia,Zhou:2012ui}.
The relation between the $S$-matrix and the on-shell reaction amplitude $T$ is given as
\begin{eqnarray}
    S_{LL'}(k)=\delta_{LL'}-\frac{i}{8\pi^2}k E_k T_{LL'}(k,k;E_k).
\end{eqnarray}
The phase shifts $\delta_{0,2}$ are complex numbers due to the presence of annihilation, 
see e.g. Refs.~\cite{Zhou:2012ui,Dai:2017ont}. Thus, we plot both the real and imaginary parts 
of the phase shifts.
%%%
As mentioned above, we only fit our amplitudes in the low-energy region, up to $T_{\rm lab}=50$~MeV. 
The fit results are shown in Fig.~\ref{Fig:PSdiffcutoff}, where the \lq data' points are taken from Ref.~\cite{Zhou:2012ui}. 
%%%%%%%%%%%%%%
Notice that, though we fit to the S-matrix elements of the PWA~\cite{Zhou:2012ui}, we plot the phase shifts and inelasticities to allow for an easy 
comparison with our previous works \cite{Dai:2017ont,Kang:2013uia}, where likewise the phase shifts were shown.  
We apply the values $\Lambda=$750, 800, 850, 900, and 950~MeV to explore the influence of the cutoff on our results. 
%%%%%%%%%%%%%%%%%%%%%%
As can be seen in Fig.~\ref{Fig:PSdiffcutoff}, all the results with different cutoffs are consistent with the PWA~\cite{Zhou:2012ui} in the energy region of $T_{\rm lab} \leq 50$~MeV, 
in general even over the larger energy region of $T_{\rm lab} \leq 100$~MeV, except for those of the $D$-waves. 

The values of the LECs for the LO and NLO potentials of our fits are listed in Table~\ref{Tab:LECs}, where we use 
the conventions of Ref.~\cite{Dai:2017ont}, that is $\tilde{C}_{{}^{2I+1\,2S+1}{L}_{J}}$ and $C_{{}^{2I+1\,2S+1}{L}_{J}}$, with the first superscript related to the isospin. 
%%%%%%%%%%%%%%%%%%%%%%%%%%%%%
\begin{table}
    \centering
 {\footnotesize
     \renewcommand{\arraystretch}{1.2}
    \begin{tabular}{|c|c|ccccc|}
        \hline
        &LO&\multicolumn{5}{c|}{NLO}  \\ \hline
        $\Lambda$\,(MeV)&850&750&800&850&900&950\\ \hline
        $\tilde{C}_{{}^{13}S_1}$(GeV$^{-2}$)  & 0.0098   & 1.3165   & 1.9390   & 2.8716   & 3.5020   & 4.5199\\
        $C_{{}^{13}S_1}$(GeV$^{-4}$)          & --       & 1.3422   & 1.0455   & 1.0649   & 0.9994   & 1.0601\\
        $C_{{}^1\epsilon_1}$(GeV$^{-4}$)      & --       & 2.4806   & 2.0745   & 1.9273   & 1.6603   & 1.3303\\
        $\tilde{C}^a_{{}^{13}S_1}$(GeV$^{-1}$)& 0.2730   &-0.6431   &-0.4821   &-0.3219   & 0.0214   & 0.5517\\
        $C^a_{{}^{13}S_1}$(GeV$^{-3}$)        & --       &-1.9000   &-0.8156   &-0.3697   &-0.1217   & 0.0363\\
        $C^a_{{}^1\epsilon_1}$(GeV$^{-3}$)    & --       &-2.0261   &-1.4201   &-1.0939   &-0.8546   &-0.6681\\
        $C^a_{{}^{13}D_1}$(GeV$^{-3}$)        & --       & 0.8432   & 0.0061   & 0.0000   & 0.0003   & 0.0000\\
        $\tilde{C}_{{}^{33}S_1}$(GeV$^{-2}$)  &-0.0531   &-0.0356   &-0.0289   &-0.0235   &-0.0304   &-0.0509\\
        $C_{{}^{33}S_1}$(GeV$^{-4}$)          &          & 0.1507   & 0.1620   & 0.1632   & 0.1727   & 0.1916\\
        $C_{{}^3\epsilon_1}$(GeV$^{-4}$)      &          & 1.1900   & 0.9196   & 0.7727   & 0.6941   & 0.6482\\
        $\tilde{C}^a_{{}^{33}S_1}$(GeV$^{-1}$)&-0.1614   & 0.0098   &-0.0041   &-0.0269   &-0.0216   & 0.0215\\
        $C^a_{{}^{33}S_1}$(GeV$^{-3}$)        & --       & 0.4299   & 0.3937   & 0.4536   & 0.4970   & 0.5050\\
        $C^a_{{}^3\epsilon_1}$(GeV$^{-3}$)    & --       &-4.7577   &-3.4012   &-2.5813   &-2.1131   &-1.8259\\
        $C^a_{{}^{33}D_1}$(GeV$^{-3}$)        & --       & 0.0001   & 0.0000   & 0.0000   & 0.0000   & 0.0000\\  \hline
    \rule[-0.3cm]{0cm}{0.7cm}{$G_{\mathrm{E}}^{0(I=0)}$}                   & 0.2807   & 0.9222   & 1.0274   & 1.1795   & 1.2406   & 1.2669   \\
  \multirow{2}{*}{$G_{\mathrm{E}}^{0(I=1)}$}  &-0.4174   &-0.0403   &-0.0057   & 0.0385   & 0.0746   & 0.1048   \\ 
                                              &+0.2979$i$&+0.5287$i$&+0.4872$i$&+0.4494$i$&+0.4134$i$&+0.3781$i$\\\hline
        $N^{\mathrm{BESIII\,2019}}_p$         & 6.0372   & 5.8976   & 5.9320   & 5.9734   & 5.9957   & 5.9902\\
        $N^{\mathrm{BESIII\,2020}}_p$         & 1.0277   & 0.9476   & 0.9495   & 0.9490   & 0.9468   & 0.9434\\
        $N^{\mathrm{BESIII\,2021}}_p$         &18.2102   &19.1115   &19.1160   &19.1184   &19.1214   &19.1258\\
        $N^{\mathrm{BaBar\,2006}}_p$          & 1.2101   & 1.1989   & 1.1987   & 1.1984   & 1.1983   & 1.1983\\
        $N^{\mathrm{BaBar\,2013}}_p$          & 1.2463   & 1.2189   & 1.2183   & 1.2173   & 1.2162   & 1.2149\\
        $N^{\mathrm{CMD-3\,2016}}_p$          & 1.3746   & 1.3274   & 1.3242   & 1.3225   & 1.3205   & 1.3169\\
        $N^{\mathrm{SND\,2014}}_n$            &10.0870   & 9.9250   &10.0633   &10.2055   &10.3328   &10.4440\\
        $N^{\mathrm{SND\,2019}}_n$            & 0.6426   & 0.6740   & 0.6769   & 0.6678   & 0.6756   & 0.6685\\
        $N^{\mathrm{SND\,2022}}_n$            & 0.6065   & 0.5967   & 0.6043   & 0.6126   & 0.6210   & 0.6295\\
        \hline
    \end{tabular}}
        \renewcommand{\arraystretch}{1.0}
    \caption{Values of parameters at LO and NLO for different cutoffs. The superscript $a$ indicates parameters that are related to the annihilation part in Eqs.~\eqref{ann}. Note that all the LECs are in units of $10^{4}$
    \cite{Dai:2017ont}. The normalization factors $N$ are discussed in more detail later. }
    \label{Tab:LECs}
\end{table}
%%%%%%%%%%%%%%%%%%%%%%%% 
Also, the other couplings, such as $G_{\mathrm{E/M}}^{0(I)}$ and the normalization factors for the 
event distribution data sets, are given, as they are fixed in the global fit.  
%%%%%
As can be observed from the black solid lines in Fig.~\ref{Fig:PSdiffcutoff}, the fit results with 
the cutoff $\Lambda=$850~MeV are overall better than those with other cutoffs. 
For instance, the one with cutoff $\Lambda=$750~MeV describes better the imaginary part of the 
$^{3}S_1$ phase shift with $I=1$ but the result for the real part is worse, whereas just the
opposite is the case for the cutoff $\Lambda=$950~MeV. 
Thus, we adopt the potentials with cutoff $\Lambda=$850~MeV for the more detailed discussion below. 
The $\chi^2$ is calculated for the $S$-matrix 
elements, in the same way as in Ref.~\cite{Dai:2017ont}: The uncertainty of 
the \lq data' (phase shifts and inelasticities) from the
PWA are estimated following Ref.~\cite{Dai:2017ont}, where we set $\Delta_{S_{LL'}}^2=0.1$. 
It is found, that the contribution from the $\bar NN$ \lq data' to the total $\chi^2$ is rather tiny. 
Indeed, the bulk of the obtained $\chi^2$ stems from $e^+e^-\to \bar NN$ 
observables like integrated and differential cross sections, 
see the discussions in the following subsection.

\begin{figure}[pt]
    \centering
    \includegraphics[width=0.9\linewidth]{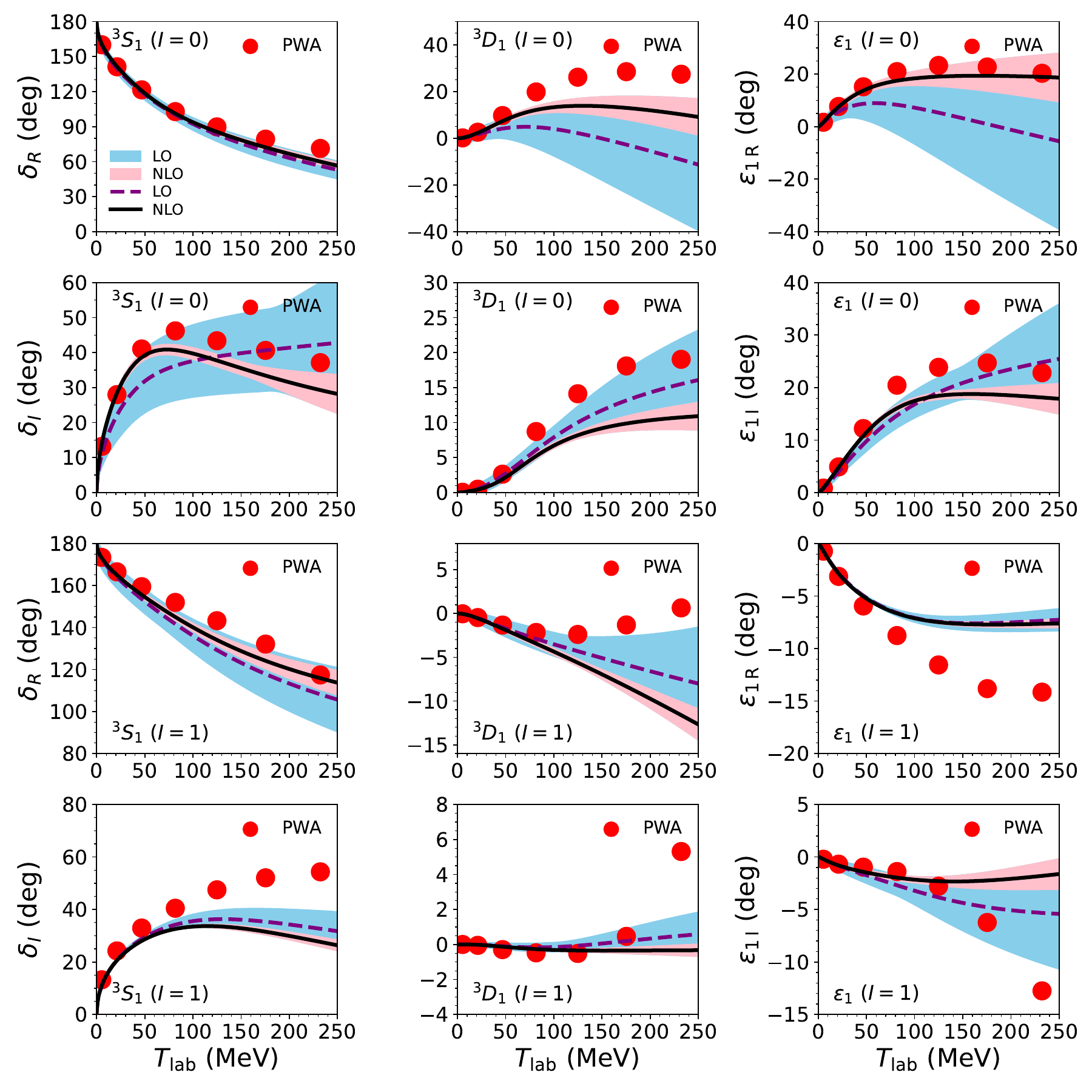}
    \caption{Real and imaginary parts of the phase shifts and inelasticities at  LO and NLO in the ${}^3S_1-{}^3D_1$ partial waves, with the isospin $I=0,1$. The red-filled circles are from the PWA~\cite{Zhou:2012ui}. The purple dashed and black solid lines are the results at LO and NLO with cutoff $\Lambda=$ 850 MeV, respectively, and the pink and sky blue bands are the corresponding uncertainties.   \label{Fig:PSdifforder} }
\end{figure}

In Fig.~\ref{Fig:PSdifforder}, we provide our results for the phase shifts and inelasticities 
at LO and NLO for $\Lambda = 850$~MeV, shown as purple dashed and black solid lines, respectively. 
%%%%
One can see that at LO the phase shifts and inelasticities of the PWA are well described 
in the very low-energy region, i.e. for $T_{\mathrm{lab}}\leq 25$~MeV, 
which includes the first two energy points of the PWA. 
In the NLO case, the `data' are well reproduced up to $T_{\mathrm{lab}}\leq 50$~MeV, 
which concerns the first three points for each set. This is in line with the expected convergence 
pattern of $\chi$EFT. As mentioned above, the scattering lengths are included in the fitting 
procedure, too. The results are listed in Table~\ref{tab:scatterlength}.

\begin{table}[ht]
    \centering
    {\footnotesize
        \renewcommand{\arraystretch}{1.2}
    \begin{tabular}{|c|cccc|}
    \hline
                             & This work: LO& This work: NLO &N$^2$LO~\cite{Kang:2013uia}&N$^3$LO~\cite{Dai:2017ont}\\
         \hline
        $a^{I=0}_{{}^{3}S_1}$(fm) & 1.43-0.72$i$  & 1.35-0.94$i$ & 1.37-0.88$i$     & 1.42-0.88$i$                          \\
        $a^{I=1}_{{}^{3}S_1}$(fm) & 0.47-0.78$i$  & 0.45-0.79$i$ & 0.44-0.91$i$    &  0.44-0.96$i$                         \\ \hline
    \end{tabular}}
        \renewcommand{\arraystretch}{1.0}
    \caption{Results of the $^3S_1$ scattering lengths (in fm) from $SU(2)$ $\chi$EFT \cite{Kang:2013uia,Dai:2017ont} and from our fit.}
    \label{tab:scatterlength}
\end{table}
The predictions by Refs.\cite{Kang:2013uia,Dai:2017ont} are set to be the \lq data'. 
The errors are set as $\Delta a=0.1$~fm, considering 
the difference between the calculations at different chiral orders \cite{Kang:2013uia,Dai:2017ont}. 
Our scattering lengths are fairly close to those predicted by $SU(2)$ $\chi$EFT.

The uncertainty is estimated following Refs.~\cite{Epelbaum:2014efa,Dai:2017ont}. The main idea is to use the expected size of higher-order corrections for the estimation of the theoretical uncertainty. The uncertainty $\Delta X^{\mathrm{NLO}}(k)$ of the NLO prediction $X^{\mathrm{NLO}}(k)$ for a given 
observable $X(k)$ can be written as~\cite{Epelbaum:2014efa}
\begin{equation} \label{eq:error}
    \Delta X^{\mathrm{NLO}}(k)=\max\left(Q^3\times|X^{\mathrm{LO}}(k)|, Q\times|X^{\mathrm{LO}}(k)-X^{\mathrm{NLO}}(k)|\right),
\end{equation}
with the parameter $Q$ defined by
\begin{equation}
    Q=\max\left(\frac{k}{\Lambda_b},\frac{M_\pi}{\Lambda_b}\right),
\end{equation}
where $k$ is the momentum in the c.m.f. and $\Lambda_b$ is the breakdown scale. Here we take $\Lambda_b$=900 MeV. Note that the quantity $X(k)$ represents either an observable such as a cross-section 
or a derived quantity, e.g., phase shifts. 
This method is expected to provide a natural and more reliable estimate of the uncertainty than relying on cutoff variations.

%%%%%%%%%%%%%%%%%%%%%%%%%%%%%%%%%%%%%%%%%%%%%%%%%%%%%%

\subsection{Fit to the $e^+e^-$ observables}
In this section, we discuss the results for the $e^+e^- \leftrightarrow \bar NN$
observables, i.e. cross sections, angular distributions, and EMFFs. 
For the relevant formulae, see Eqs.~(\ref{eq:vertex}-\ref{eq:sigma}). 
Note that these data and the ones for $\bar{N}N$ scattering are fitted simultaneously. 
For $e^+e^-\to \bar{p}p$, the data sets of cross sections are taken form ADONE73~\cite{Castellano:1973wh}, Fenice~\cite{Antonelli:1993vz,Antonelli:1994kq,Antonelli:1998fv}, DM1~\cite{Delcourt:1979ed}, DM2~\cite{Bisello:1983at}, BaBar~\cite{BaBar:2005pon,BaBar:2013ves}, CMD-3~\cite{CMD-3:2015fvi,CMD-3:2018kql}, BESIII~\cite{BES:2005lpy,BESIII:2019hdp,BESIII:2019tgo,BESIII:2021rqk}. 
Data for $e^+e^-\to \bar{n}n$ are taken from Fenice~\cite{Antonelli:1993vz,Antonelli:1998fv}, SND~\cite{Achasov:2014ncd,Druzhinin:2019gpo,SND:2022wdb} and BESIII~\cite{BESIII:2021tbq,BESIII:2022rrg}. 
The  $\bar{p}p\to e^+ e^-$ cross sections are taken from PS170~\cite{Bardin:1994am}. 
However, it should be stressed that, since some old data sets have significant errors, 
we only include data published after 2005 in the actual fitting procedure,
that is, the data from BaBar~\cite{BaBar:2005pon,BaBar:2013ves}, CMD-3~\cite{CMD-3:2015fvi,CMD-3:2018kql}, and BESIII~\cite{BESIII:2019hdp,BESIII:2019tgo,BESIII:2021rqk} 
for $e^+e^-\to \bar{p}p$, and SND~\cite{Achasov:2014ncd,Druzhinin:2019gpo,SND:2022wdb} and BESIII~\cite{BESIII:2021tbq,BESIII:2022rrg} for $e^+e^-\to \bar{n}n$. 
An exception is made for the data by PS170~\cite{Bardin:1994am}
since it is the only experimental information for the reaction $\bar{p}p\to e^+ e^-$.

\begin{figure}[ht]
\centering
\includegraphics[width=0.8\linewidth]{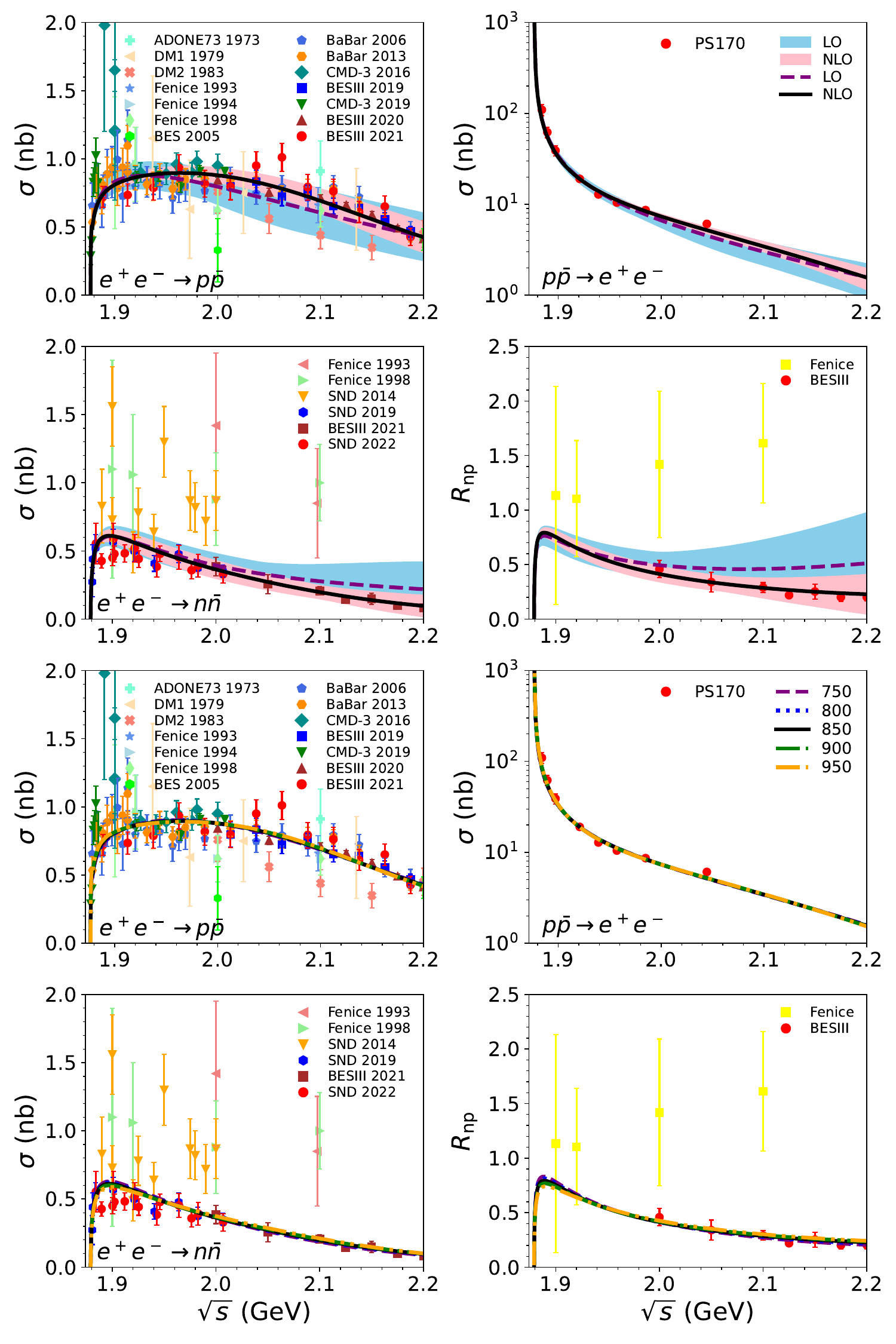}
\caption{The comparison between our fit results and the experimental data sets of cross sections, including the processes of $e^+e^-\to\bar{p}p$, $\bar{n}n$, and $\bar{p}p\to e^+ e^-$. The experimental data sets are taken from ADONE73~\cite{Castellano:1973wh}, Fenice~\cite{Antonelli:1993vz,Antonelli:1994kq,Antonelli:1998fv}, DM1~\cite{Delcourt:1979ed}, DM2~\cite{Bisello:1983at}, BaBar~\cite{BaBar:2005pon,BaBar:2013ves}, CMD-3~\cite{CMD-3:2015fvi,CMD-3:2018kql}, BESIII~\cite{BES:2005lpy,BESIII:2019hdp,BESIII:2019tgo,BESIII:2021rqk}, 
SND~\cite{Achasov:2014ncd,Druzhinin:2019gpo,SND:2022wdb}, and PS170 \cite{Bardin:1994am}.
    \label{fig:sigma1} }
\end{figure}

Our fits to the cross sections for $e^+e^-\to \bar{p}p$, $\bar{n}n$, and 
$\bar{p}p\to e^+ e^-$ are summarized in Fig.~\ref{fig:sigma1}. 
The four graphs at the top are the fit results for LO (purple dashed) and NLO (black solid) 
for the cutoff $\Lambda=850$~MeV, with the corresponding error bands in the colors sky-blue and pink, respectively. 
The four graphs at the bottom are NLO results with different cutoffs: The purple dashed, 
blue dotted, black solid, green dash-dotted, and orange dash-dot-dotted lines are for 
cutoffs 750, 800, 850, 900, and 950~MeV, respectively.
As can be seen, the LO results are consistent with the data up to roughly $2$~GeV, 
while the NLO results agree with the experiments rather 
well over the whole considered energy region, for all cutoffs.  

Obviously, the $e^+e^-\to \bar NN$ cross sections rise very quickly 
from the $\bar NN$ thresholds, see Fig.~\ref{fig:sigma1}. Then, for $e^+e^-\to \bar{p}p$, the cross section remains unchanged up to roughly 2 GeV, and eventually starts to decrease. Regarding $\bar{p}p \to e^+e^-$, the inverse process of the former, the cross-section decreases rapidly with energy. The difference in the behavior can be easily understood from the relation between
the reaction cross sections, $\sigma_{\bar{p}p\to e^+ e^-}
\simeq (k_e^2/k_{N}^2)\, \sigma_{e^+ e^-\to \bar{p}p}$ which follows from 
time reversal invariance~\cite{Haidenbauer:2014kja}. 
For the cross section of the process $e^+e^-\to \bar{n}n$, see the left side graphs in the second and fourth rows of Fig.~\ref{fig:sigma1}, there is also a strong rise near the threshold, but the situation for the proton and neutron cases is a bit different: the one for the neutron starts to decrease rapidly at 1.9 GeV. As discussed in Ref.~\cite{Yang:2022qoy}, this reveals that the oscillation of the so-called subtracted form factors (SFFs) 
of the neutron and proton are different. 
The ratio of the cross sections, $R_{np}=\sigma(e^+e^-\to\bar{n}n)/\sigma(e^+e^-\to\bar{p}p)$, are essential to refine the analysis. As can be found, ours fit the data well. See the last graph in the second and fourth rows. Nevertheless, in the low-energy region, the statistics of the data are poor.  
It would be rather helpful to perform more experiments in the energy region closer to the 
$\bar NN$ thresholds.

%%%%%%%%%%%%%%%%%%%%
\begin{figure}[ht]
    \centering
    \includegraphics[width=0.99\linewidth]{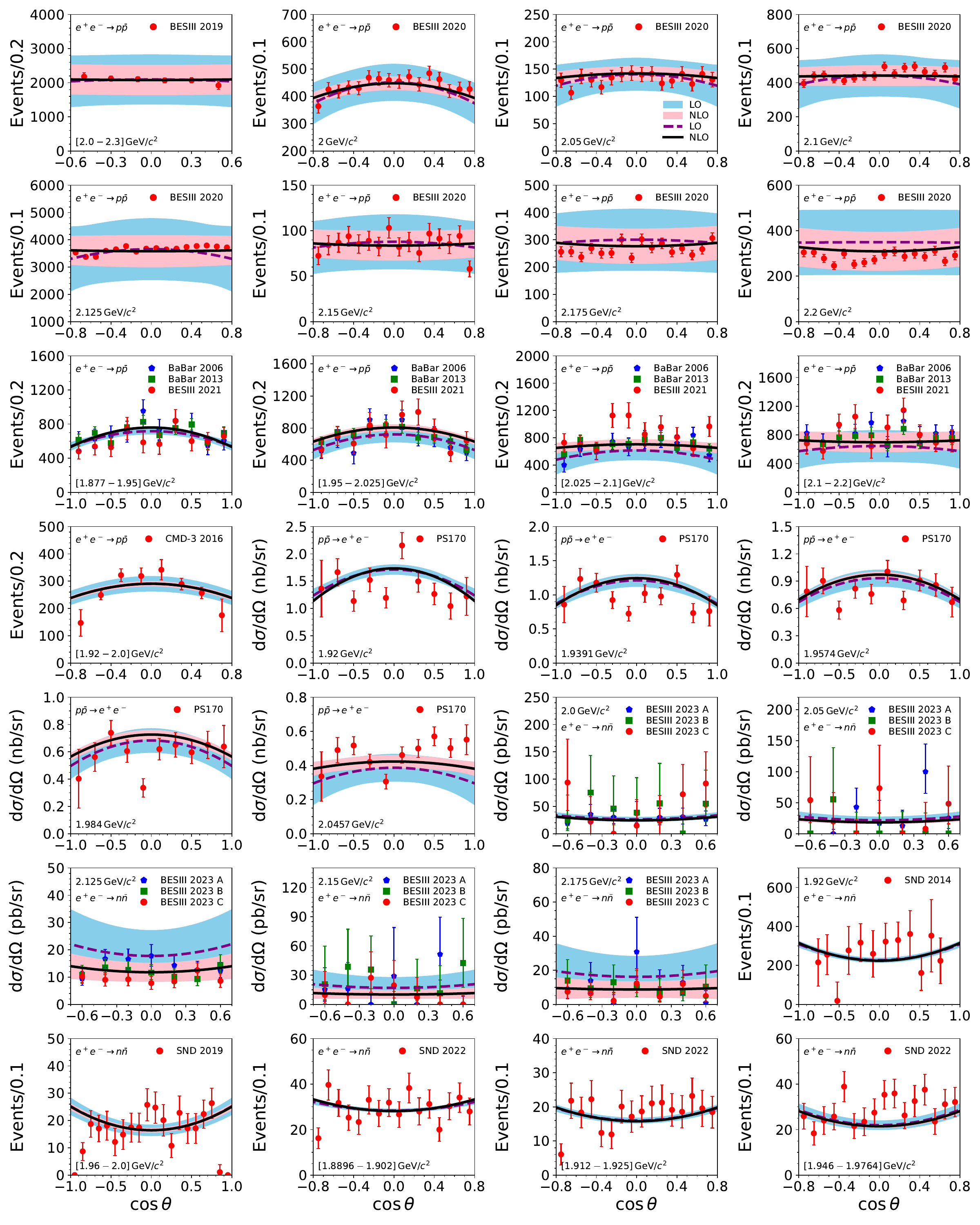}
    \caption{Our fit to the angular distributions of the processes of $e^+e^-\to \bar{p}p$, $\bar{n}n$, and $\bar{p}p\to e^+ e^-$. The experimental data are taken from BESIII \cite{BESIII:2019tgo,BESIII:2019hdp,BESIII:2021rqk,BESIII:2022rrg}, BaBar \cite{BaBar:2005pon,BaBar:2013ves}, CMD-3 \cite{CMD-3:2018kql}, SND~\cite{Achasov:2014ncd,Druzhinin:2019gpo,SND:2022wdb} and PS170~\cite{Bardin:1994am}.}
    \label{fig:dsigma1}
\end{figure}
%%%%%%%%%%%%%%%%%%%%%
%
Experimental results for the angular distributions of $e^+e^-\to \bar{p}p$ are available for the energy intervals of 
1.877-1.950 GeV$/c^2$, 1.950-2.025 GeV$/c^2$, 2.025-2.100 GeV$/c^2$, 2.100-2.200 GeV$/c^2$, 
1.920-2.000 GeV$/c^2$, 2.000-2.300 GeV$/c^2$, 1.920-2.000 GeV$/c^2$, 
and at 2.000, 2.050, 2.100, 2.125, 2.150, 2.175, 2.2 GeV$/c^2$. 
The data are taken from the works of the BESIII \cite{BESIII:2019tgo,BESIII:2019hdp,BESIII:2021rqk},  
BaBar \cite{BaBar:2005pon,BaBar:2013ves}, and CMD-3 \cite{CMD-3:2018kql} collaborations.
%%%
In Ref.~\cite{Bardin:1994am}, angular distributions for $\bar{p}p\to e^+ e^-$ 
can be found at the energies $\sqrt{s}$=1.9200, 1.9391, 1.9574, 1.9840, and 2.0457 GeV. 
Angular distributions for $e^+e^-\to \bar{n}n$ were measured at energy intervals of 1.960-2.000, 1.8896-1.9020, 1.9120-1.9250, and 1.9460-1.9764 GeV$/c^2$, 
and at the energies $\sqrt{s}$=1.920, 2.000, 2.050, 2.125, 2.150, end 2.175 GeV$/c^2$. 
They are taken from the SND~\cite{Achasov:2014ncd,Druzhinin:2019gpo,SND:2022wdb} and BESIII~\cite{BESIII:2022rrg} experiments. 

A visual comparison between our results and the data is provided in Fig.~\ref{fig:dsigma1}. 
%%%%%%%%%%%%
The purple dashed and black solid lines are the results at LO and NLO, respectively. Correspondingly, the sky-blue and pink bands are the uncertainties of LO and NLO, 
calculated from Eq.~(\ref{eq:error}).   
Notice that for each data set of the event distribution we apply one constant normalization factor for all energy values.  
For instance, for all the data points of BESIII in the year 2020 \cite{BESIII:2019hdp}, as shown in the first two rows, 
the normalization factor is labeled as $N^{\mathrm{BESIII\,2020}}_p$.

As can be seen from Fig.~\ref{fig:dsigma1}, our fit is of high quality, and this confirms the reliability of our analysis.
The graphs of the first four rows and the first two graphs in the fifth row are our results for the angular distributions for $e^+e^-\to \bar{p}p$ 
and $\bar{p}p\to e^+ e^-$, where the last five graphs in the indicated places are that for the latter reaction. 
The results for $e^+e^-\to \bar{p}p$ are better than that of the $\bar{p}p\to e^+ e^-$. This is partly caused by the lower statistics 
of the latter data. Nevertheless, the $\bar{p}p\to e^+ e^-$ differential cross-section data have no normalization factors, and ours are 
consistent with them within the errors. 
%%%
The graphs in the three bottom rows (except for the first two graphs) are our fits to the data sets of the angular distributions of $e^+e^-\to \bar{n}n$. 
The data for the differential cross section have significant uncertainties except for the one at the energy point $\sqrt{s}=2.125$~GeV. 
Our results agree well with these data points, too. 
The data on the angular distributions have more minor errors, but the fit quality is not as good as for the other two processes, 
$e^+e^-\to \bar{p}p$ and $\bar{p}p\to e^+ e^-$. Nevertheless, our results are still compatible with the data sets except for a few 
points near $\cos \theta=\pm1$. Notice that the measurements are difficult to perform close to $\cos \theta=\pm1$. % from the experimental point of view. 

%%%%%------------------------------------------
\begin{table}[htbp]
	\centering
	{\footnotesize
     \renewcommand{\arraystretch}{1.3}
		\begin{tabular}{|c|cc|cccccc|}
			\hline
			\rule[-0.3cm]{0cm}{8mm} & \multicolumn{2}{c|}{LO} & \multicolumn{6}{c|}{NLO}\\
			&$N$&$\chi^2/N$&$N$&\multicolumn{5}{c|}{$\chi^2/N$} \\
			\hline
			$\Lambda$ (MeV)   & & 850& &750 & 800 & 850 & 900 & 950\\
			\hline
			Cross Section                                                               & 105 & 1.59 & 154 & 1.70 & 1.65 & 1.58 & 1.53 & 1.48\\
			Differential cross section                                                  & 221 & 1.31 & 477 & 1.59 & 1.57 & 1.53 & 1.49 & 1.47\\
			$R_{np}$                                                                    &   1 & 0.20 &   7 & 0.38 & 0.62 & 0.99 & 1.41 & 1.76\\
			%Effective EMFFs                                                             &  88 & 1.89 & 135 & 1.60 & 1.60 & 1.59 & 1.60 & 1.66\\
			$|G_{\mathrm{E}}/G_{\mathrm{M}}|$, $|G_{\mathrm{E}}|$ and $|G_{\mathrm{M}}|$&  13 & 0.54 &  44 & 1.74 & 1.70 & 1.60 & 1.42 & 1.22\\
			% $G_{\mathrm{osc}}$                                                          &  33 & 2.59 &  71 & 1.66 & 1.65 & 1.66 & 1.71 & 1.85\\
			Phase shift                                                                 &  24 & 0.008&  36 & 0.003& 0.004& 0.004& 0.005&0.006\\
			Scattering length                                                           &   4 & 1.41 &   4 & 0.86 & 0.92 & 0.93 & 0.87 & 0.84\\ \hline
			total                                                  &  368 & 1.28 &  722 & 1.53 & 1.50 & 1.46 & 1.42 & 1.38 
 \\
			\hline
	\end{tabular}}
     \renewcommand{\arraystretch}{1.0}
	\caption{The $\chi^2/N$ of cross section, differential cross section, $|G_{\mathrm{E}}/G_{\mathrm{M}}|$, $|G_{\mathrm{E}}|$, $|G_{\mathrm{M}}|$ and $R_{np}$ for the NLO and LO fits. 
 In the last row we list the total $\chi^2/N$ to provide an overview 
 of the fit quality. 
 \label{Tab:chi2}}
\end{table}
%%%%%------------------------------------------
In order to provide a quantitative overview of the quality of our results, 
we summarize the $\chi^2$ values of the fits for each cutoff $\Lambda$ 
in Table \ref{Tab:chi2}. 
As can be seen, most of the available data are for total and differential 
cross sections.
For each kind of data set, our fits yield a $\chi^{2}/N$ around one, 
where $N$ is the number of data points,  
while the contributions from phase shifts (by fitting to the S-matrix elements) and scattering lengths are tiny. 
Finally, the total $\chi^2/N$ can be found in the last row of Table \ref{Tab:chi2}. The value for the fit with cutoff $\Lambda=850$ MeV is 
about 1.46. This indicates that our fit is of high quality and it can be used 
to extract the individual EMFFs reliably.  

\subsection{Extracting individual EMFFs}
\begin{figure}[ht]
    \centering
    \includegraphics[width=0.8\linewidth, height=0.23\textheight]{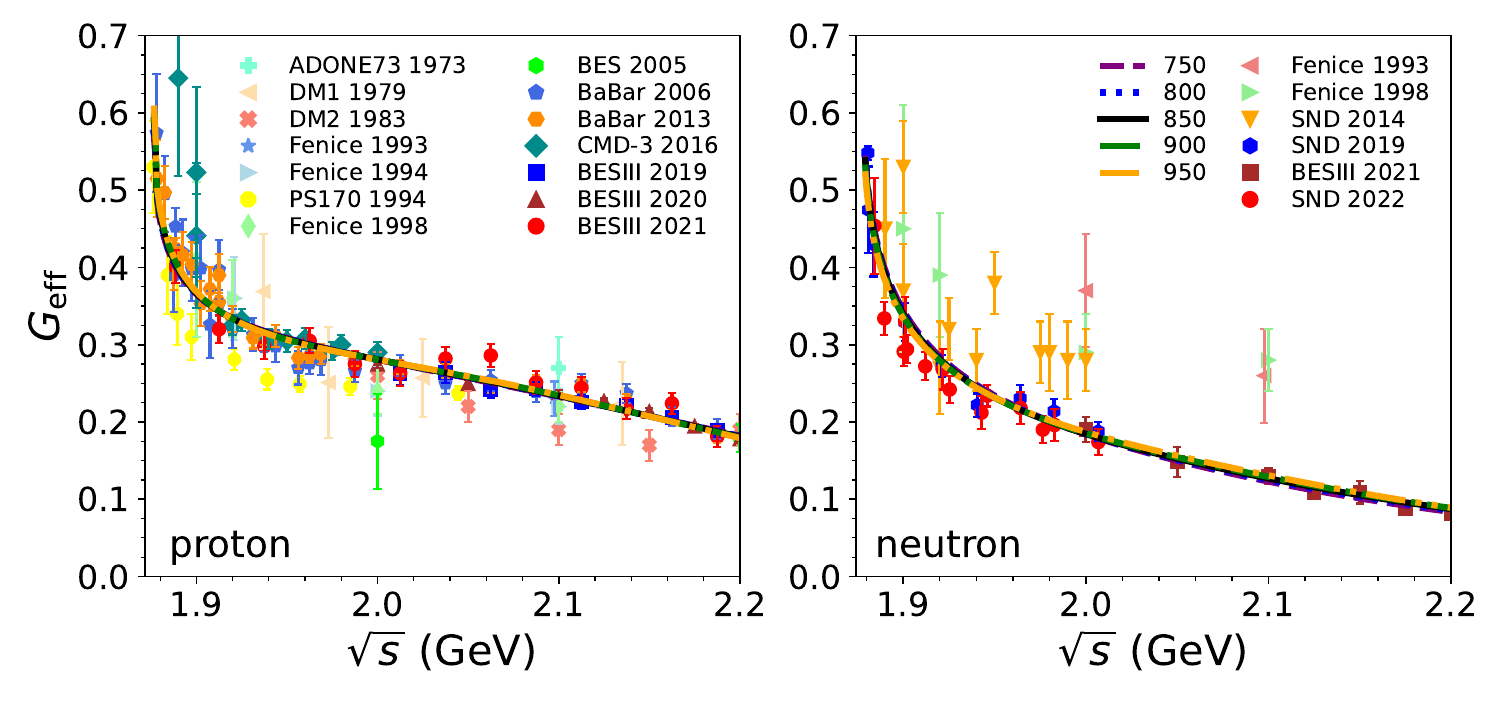}
    \caption{The comparison between our results and the experimental datasets of effective EMFFs. The experimental data sets are taken from ADONE73~\cite{Castellano:1973wh}, Fenice~\cite{Antonelli:1993vz,Antonelli:1994kq,Antonelli:1998fv}, DM1~\cite{Delcourt:1979ed}, DM2~\cite{Bisello:1983at}, BaBar~\cite{BaBar:2005pon,BaBar:2013ves}, CMD-3~\cite{CMD-3:2015fvi}, BESIII~\cite{BES:2005lpy,BESIII:2019hdp,BESIII:2019tgo,BESIII:2021rqk,BESIII:2021tbq}, SND~\cite{Achasov:2014ncd,Druzhinin:2019gpo,SND:2022wdb}, PS170~\cite{Bardin:1994am}.  
    \label{Fig:effective} }
    \end{figure}

%%%%%%%%
The effective EMFF, $|G_{\mathrm{eff}}|$, is basically a parameterization 
of the total cross section, cf. Eq.~(\ref{eq:Geff}), 
which is provided by many experimental groups. 
We present pertinent results in Fig.~\ref{Fig:effective}. 
%%%%%%%%%
%%%%%%%%%%%%%%%%%%%%%%%%%%
%%%%%%%%%%%%%%%%%%%%%%%%%
The graph on the left side shows the effective EMFF of the proton, and the one on the right side is that of the neutron. 
The purple dashed, blue dotted, black solid, green dash-dotted, and orange dash-dot-dotted lines are our results at NLO with cutoffs $\Lambda=$750, 800, 850, 900, 950 MeV, respectively. Obviously, these lines overlap with each other, 
and for all cutoffs an excellent description
of the effective EMFFs  is obtained.
The effective EMFFs of the proton and the neutron exhibit similarities and
differences. Both effective EMFFs fall off rapidly for energies near 
the $\bar NN$ threshold, and then much more slowly with increasing energy. 
However, there is a difference in the magnitude, with the effective form 
factor of the proton being noticeably larger than that of the neutron. 
Indeed, as will be discussed in more detail in appendix \ref{app:EMFFs}, 
the dipole functions of the proton and neutron are related by 
$\mathcal{A}^p\simeq 2 \mathcal{A}^n$, which reflects the observation discussed above. Moreover, there is an extra energy factor $1/[1+s/m_a^2]$ for the proton. 
The dynamical reason of the difference on the dipole functions may be that the proton is electrically charged and the neutron is electrically neutral. 
As is argued in the Supplement of Ref.~\cite{Yang:2022qoy}, the dipole function contributes mainly to the effective EMFFs.         
It would be rather helpful if experiments give more information in the energy region closer to the nucleon thresholds.
In appendix \ref{app:EMFFs} we also discuss the oscillation of the SFFs in the energy region below 2.2 GeV. 
The results are similar to our earlier paper \cite{Yang:2022qoy}. 

%%%%%%%%%%%%%%
\begin{figure}[ht]
    \centering
    \includegraphics[width=0.99\linewidth]{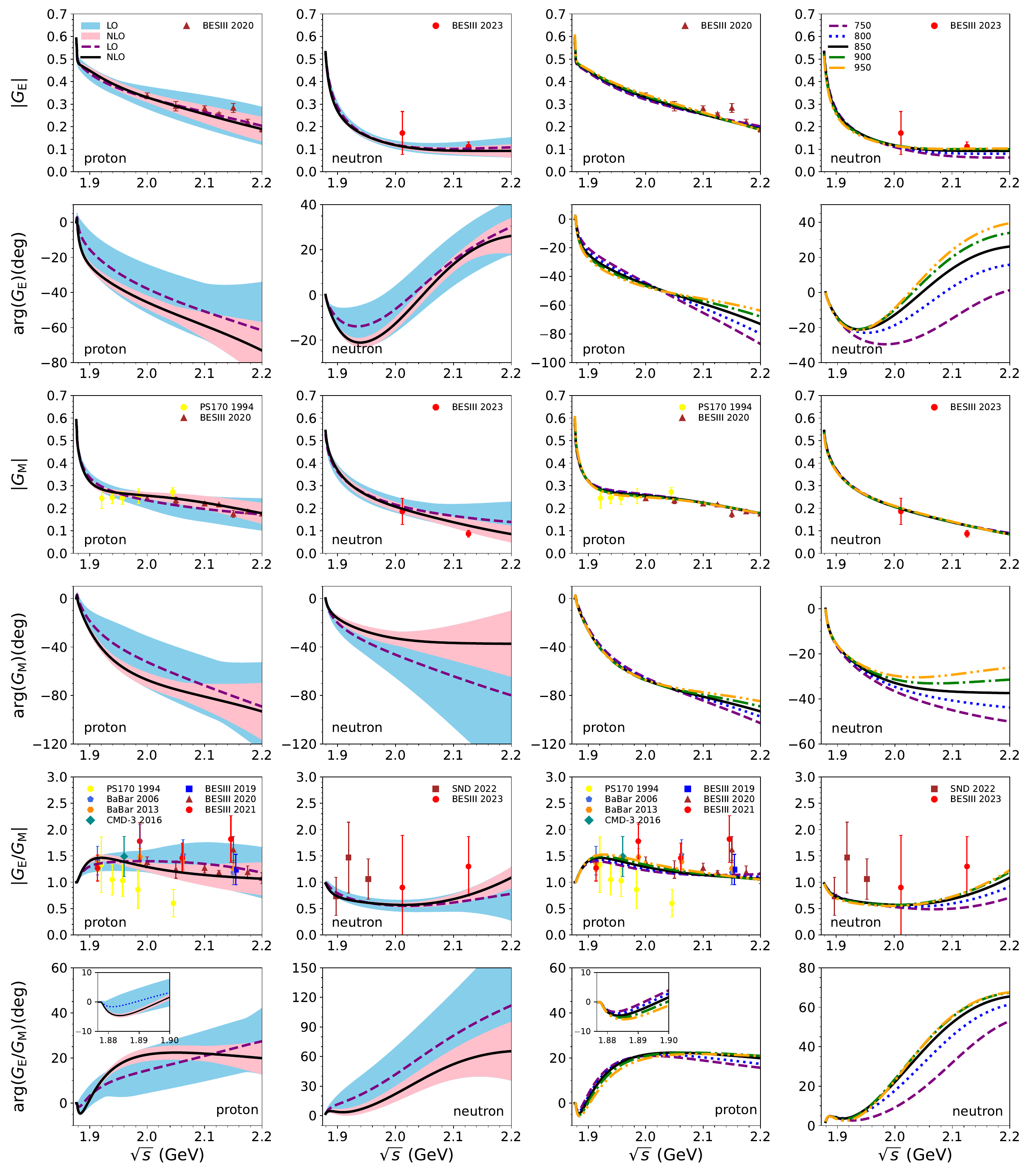}
    \caption{Predictions for the individual EMFFs, $G_{\mathrm{E}}$, $G_{\mathrm{M}}$, and the ratios $G_{\mathrm{E}}/G_{\mathrm{M}}$, including the modulus and the phases. 
    The data shown in the graphs are taken from BaBar~\cite{BaBar:2005pon,BaBar:2013ves}, CMD-3~\cite{CMD-3:2015fvi}, BESIII~\cite{BESIII:2019hdp,BESIII:2019tgo,BESIII:2021rqk,BESIII:2022rrg}, SND~\cite{SND:2022wdb}, PS170~\cite{Bardin:1994am}. The results in the left two columns are for the
    cutoff $\Lambda=850$ MeV.
    \label{Fig:gme} }
    \end{figure}
    %%%%%%%%%%%%%%
From the reaction amplitude, one can also extract the individual EMFFs. Results
for the individual EMFFs, $G_{\mathrm{E}}$  
and $G_{\mathrm{M}}$, and their ratio $G_{\mathrm{E}}/G_{\mathrm{M}}$, including both the modulus and the phases, are shown in Fig.~\ref{Fig:gme}. 
The graphs in the first two columns are our results at LO and NLO, with cutoff $\Lambda=850$~MeV, where the 
purple dashed and black solid lines are for LO and NLO, with the corresponding error bands sky blue and pink, respectively.
The graphs in the last two columns are the results at NLO with different cutoffs, $\Lambda=750, 800, 850, 900, 950$~MeV, corresponding to
the purple dashed, blue dotted, black solid, green dash-dotted, and orange dash-dot-dotted lines, respectively.  
%.  
There are only several published points for the modulus of the individual EMFFs, and our results are consistent with them.

The electric and magnetic form factors exhibit a similar behavior as the effective EMFFs, i.e. all of them decrease 
rapidly around the thresholds and more slowly for higher energies. 
Regarding the proton, the electric form factor decreases more rapidly with increasing $\sqrt{s}$ above 2~GeV, compared with the magnetic form factor.
In case of the neutron it is the opposite. 
As shown in the graphs of the third row of Fig.~\ref{Fig:gme}, the ratio of $|G_{\mathrm{E}}/G_{\mathrm{M}}|$ 
of the proton increases above one 
immediately after the threshold and then decreases with increasing energy $\sqrt{s}$, while the ratio of the neutron form factors decreases around the threshold and then increases as the energy increases.  
For the proton, the turning point of the ratio is at around 1.9 GeV, while that of the neutron is at 2.0 GeV. 
Both of the ratios start from one at the threshold by definition. But, interestingly,  
it looks as if both of them will go back to one  at higher energies, 
implying that $|G_{\mathrm{E}}|$ would become equal to 
$|G_{\mathrm{M}}|$ again in the high energy region. 
This can be checked by future experiments. 

The phases are shown in the second, fourth, and sixth rows in Fig.~\ref{Fig:gme}. Note that there can be an overall phase factor for the individual EMFFs, which can not be observed. Therefore, we set all the phases to be zero at the relevant thresholds. The variations of the phases with the energies are much more different from that of the modulus. As can be seen, the phases of all the individual EMFFs decrease monotonously as the energy increases, except for that of the electric form factors of the neutron, which
decreases first and then increases. It may indicate the difference between the charge of the valence quarks for the proton and the neutron.  
Interestingly, there is a peculiar behavior of the phases of 
$G_{\mathrm{E}}/G_{\mathrm{M}}$ very close to the thresholds, see 
the enlarged graphs in the last row of Fig.~\ref{Fig:gme}
(and also Fig.~8 in \cite{Haidenbauer:2014kja}).
Obviously the phase between the electric and the magnetic form factors varies
significantly in the low-energy region, i.e. over the first 100~MeV or so.
This is not caused by the mass difference between the proton and neutron which is ignored since the $\bar{N}N$ scattering amplitudes are evaluated in the
isospin basis.
Actually, the minimum of the phase is at roughly 1.883~GeV, i.e. above the thresholds of $\bar{p}p$ or $\bar{n}n$.  
We note that the electric and magnetic form factors have almost no phase difference without FSI. Also, the phase difference is much smaller for the LO results, where the D-wave contributions are smaller\footnote{In
Ref.~\cite{Guo:2024pti} a similar behavior is observed in the LO calculation of the EMFFs of $\Lambda_c$. However, the one in Ref.~\cite{Guo:2024pti} is more flat. This is so because in the $\Lambda_c^+\bar {\Lambda_c}^-$ case 
there is no D-wave contribution at LO.}.   
The strong variation for the NLO FSI reveals that there is a remarkable
sensitivity, reflected in the properties of the EMFFs near the threshold. 
This is interesting and deserves 
further study through both theory and experiment.   

 %%%%%%%%%%%%%%%%%%%%%%%%%%%%%%%%%%%%%%%%%%%%%%%%%%%%%%%%%%

\section{Conclusions}
\label{Sec:IV}
In this paper we evaluated the individual EMFFs of the proton and neutron in the 
processes $e^+e^-\to \bar{p}p$ and $e^+ e^-\to \bar{n}n$. 
The final-state interaction between antinucleon and nucleon is taken 
into account. The latter is based on a $\bar{N}N$ scattering amplitude
generated from a $\bar{N}N$ potential derived within $SU(3)$ 
$\chi$EFT up to NLO.
It is included in the calculation of the integrated and differential cross sections of the processes $e^+e^-\to \bar{N}N$ within the DWBA. 

An excellent description of the available data on the reaction $e^+e^-\to \bar{N}N$ up to 2.2~GeV is achieved. Specifically, our calculations also reproduce the strong enhancement of the cross sections near the $\bar NN$ thresholds observed in both processes. The individual EMFFs, $G_{\mathrm{E}}$ and $G_{\mathrm{M}}$, of the proton and neutron, and their ratio $G_{\mathrm{E}}/G_{\mathrm{M}}$, including the modulus and the phases, are predicted. 
It turned out that the phases of the electric form factors of the proton and the neutron are quite different. 
Interestingly, it is found that the relation $|G_{\mathrm{E}}|=|G_{\mathrm{M}}|$,
strictly valid at the $\bar NN$ threshold, is eventually restored in the 
higher-energy region. 

More accurate measurements near the threshold, hopefully performed in the near future, will be essential to get more precise constraints on electric and magnetic form factors of the nucleons and are important to refine our analysis. This will be helpful to understand the properties of the nucleons as well as the strong interaction.

\section*{Acknowledgements}
\label{Sec:V}
This work is supported by the National Natural Science Foundation of China (NSFC) with Grants No.12322502, 12335002, 11805059, 11675051, Joint Large Scale Scientific Facility Funds of the NSFC and Chinese Academy of Sciences (CAS) under Contract No.U1932110, and Fundamental Research Funds for the central universities. 
It was further supported by Deutsche Forschungsgemeinschaft (DFG) and NSFC through funds provided to the Sino-German CRC 110 ``Symmetries and the Emergence of Structure in QCD" (NSFC Grant No.~11621131001, DFG Grant No.~TRR110).
The work of UGM was supported in part by the CAS President's International Fellowship Initiative (PIFI) (Grant No.~2018DM0034).

\appendix

%%%%%%%%%%%%%%%%%%%%%%%%%%%%%%%%%%%%%%%%%%%%%%%%%%%%%%%

\section{Oscillation of the subtracted form factors}\label{app:EMFFs}
% %%%%%%
The effective EMFFs $G^N_{\mathrm{eff}}$ ($=G^N_{\E}=G^N_{\M}$ ) \cite{Castellano:1973wh,Delcourt:1979ed} 
have been been published for some experiments. 
They can be extracted from the integrated cross section directly
\begin{align}
|G_{\mathrm{eff}}(s)|=\sqrt{\frac{\sigma_{e^+e^-\to \bar{N}N}(s)}{\frac{4\pi\alpha^2\beta}{3s}C(s)[1+\frac{2M_N^2}{s}]}}. \label{eq:Geff} 
\end{align}
An interesting phenomenon was discovered in the analysis of the experiment 
concerning these effective EMFFs: 
they show an oscillatory behavior once the dipole contributions is subtracted. 
Moreover, Ref.~\cite{BESIII:2021tbq} found a phase difference between the oscillation of the 
subtracted form factors (SFFs) of the proton and neutron. 

%%%%%%%%%%
\begin{figure}[ht]
    \centering
    \includegraphics[width=0.8\linewidth]{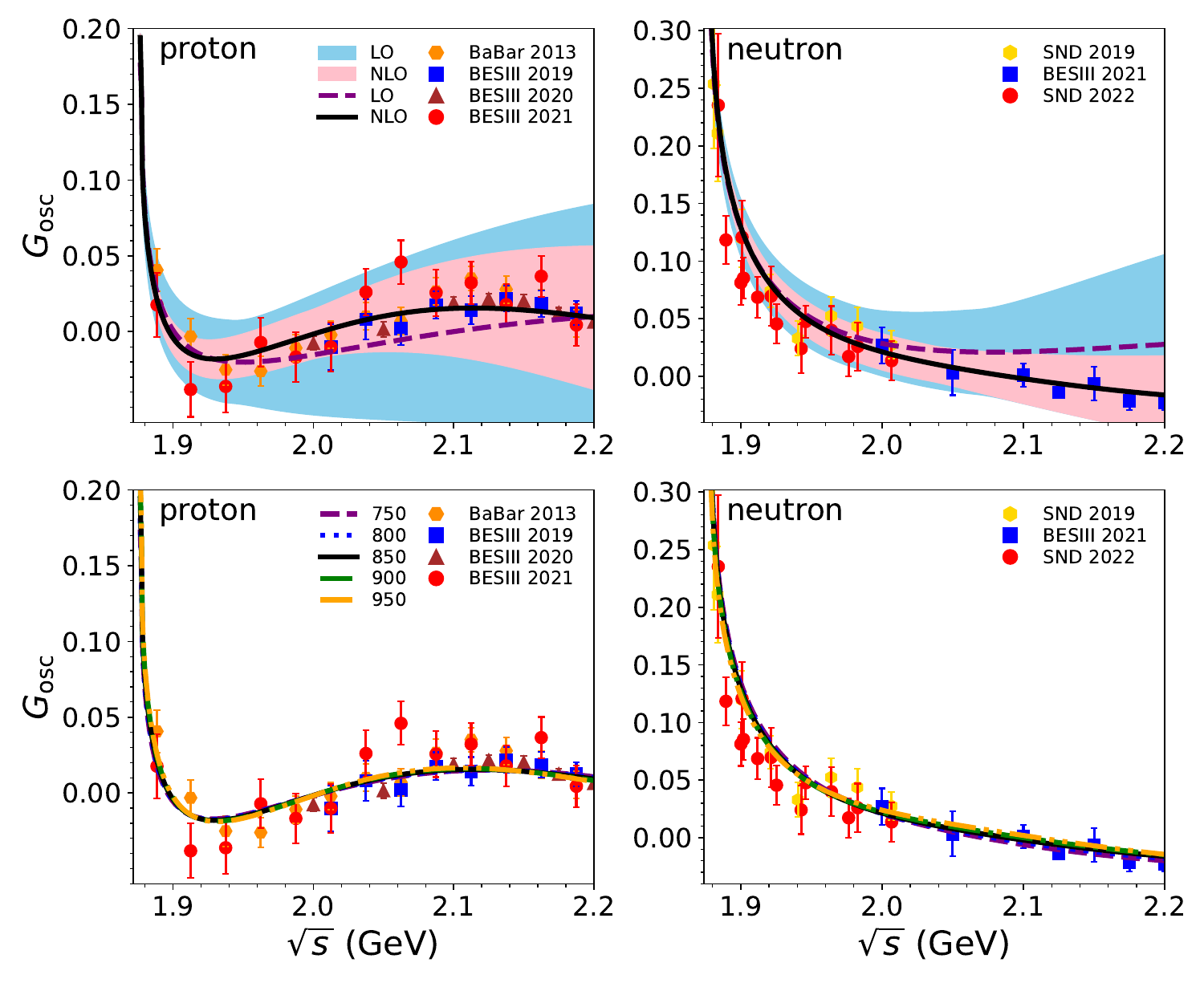}
    \caption{Comparison between our predictions and the oscillations of the SFFs deduced from 
the experiments. The latter results
are from BaBar~\cite{BaBar:2005pon,BaBar:2013ves}and BESIII~\cite{BESIII:2019hdp,BESIII:2021rqk}, 
SND~\cite{Druzhinin:2019gpo,SND:2022wdb}. 
The purple dashed and black solid lines are our results at LO and NLO, with the 
corresponding error bands sky blue and pink, respectively. 
The cutoff is chosen to be $\Lambda=850$ MeV.
    \label{Fig:gosc}}
    \end{figure}
%%%%%%%%%%%
The SFFs of the nucleons are defined as the difference between the effective EMFFs and the 
dipole contribution~\cite{Bianconi:2015owa,BESIII:2021rqk,BESIII:2021tbq}, 
\begin{equation}
    G_{\mathrm{osc}}(s)=|G_{\mathrm{eff}}|-G_D(s)\,, \label{eq:G;osc}
\end{equation}
where $G_D$ is the dipole function given as \cite{Bianconi:2015owa,BESIII:2021rqk}
\begin{eqnarray}
    G_D^p(s)&=&\frac{\mathcal{A}_p}{(1+s/m_a^2)[1-s/q_0^2]^2} \,, \nonumber \\
    G_D^n(s)&=&\frac{\mathcal{A}_n}{[1-s/q_0^2]^2} \,. \nonumber
\end{eqnarray}
Here, the parameters are $\mathcal{A}_p$=7.7, $\mathcal{A}_n$=3.5$\pm$0.1, $m_a^2$=14.8 ($\mathrm{GeV}/c$)$^2$ and $q_0^2$=0.71 ($\mathrm{GeV}/c$)$^2$. 
Our predictions for the SFFs are shown in Fig.~\ref{Fig:gosc}. They are simply evaluated via 
Eq.~\eqref{eq:G;osc}, with the $G_{\mathrm{eff}}$ obtained by Eq.~\eqref{eq:Geff}. 
Note that in the present analysis the FSI near the thresholds is considered\footnote{One should be aware that the FSI~\cite{Dai:2014zta,Dai:2016ytz} can change the Born amplitudes dramatically especially in the energy region close to the threshold.} and our results are compatible with the experimental data in that region.  
This is not the case for some other pertinent studies in the literature.

As can be seen, when $\sqrt{s}$ is close to the $\bar NN$ threshold, the oscillation 
pattern of both proton and neutron disappears gradually.
Besides, it looks like that there is no phase difference between the SFFs ($G_{\mathrm{osc}}$)
of the proton and that of the neutron. These are consistent with the discussion in our earlier
paper \cite{Yang:2022qoy}, where the overdamped oscillators dominate in the low-energy region, 
and the oscillation is not so straightforward as that in the high-energy region 
above 2~GeV, which is dominated by the underdamped oscillators. 
Also, there is no phase difference between the overdamped oscillators.  
Of course, this conclusion remains to be checked by more accurate measurements near the threshold 
in future experiments. 

%%%%%%%%%%%%%%%%%%%%%%%%%%%%%%%%%%%%%%%%%%%%%%%%%%%%%%%%%%

\section{The results of the TBE potential}\label{app:TBE}
In Sec.~\ref{sec:TBE}, the calculation of the football diagram is described. In this section, 
the potentials due to the triangle, planar box, and cross box diagrams will be given. 
The triangle diagrams refer to the left triangle and right triangle diagrams. For the left 
triangle diagrams in the physical basis, one has
\begin{eqnarray}
V^{\mathrm{Left Triangle}}_{\bar{\n}\n\to\bar{\n}\n}\!\!&=&\!\!V^{\mathrm{Left Triangle}}_{\bar{\p}\p\to\bar{\p}\p}\nonumber\\
        \!\!&=&\!\!-\frac{f_{NN\pi}^2}{8f_0^2}\!\!\int\!\!\frac{\dd^3\bm{l}_1}{(2\pi)^3}\frac{(\bm{l}_1^2\!-\!\bm{q}^2)\!\langle\lambda'_2|\lambda_2\rangle\!\langle\lambda'_1|\lambda_1\rangle}{\omega_{+,\pi}\omega_{-,\pi}\!(\omega_{+,\pi}\!+\!\omega_{-,\pi})}\!-\!\frac{f_{\Lambda NK}^2}{8f_0^2}\!\!\int\!\!\frac{\dd^3\bm{l}_1}{(2\pi)^3}\frac{(\bm{l}_1^2\!-\!\bm{q}^2)\!\langle\lambda'_2|\lambda_2\rangle\!\langle\lambda'_1|\lambda_1\rangle}{\omega_{+,K}\omega_{-,K}\!(\omega_{+,K}\!+\!\omega_{-,K})}\nonumber\\
        \!\!&&\!\!-\frac{f_{\Sigma NK}^2}{4f_0^2}\!\!\int\!\!\frac{\dd^3\bm{l}_1}{(2\pi)^3}\frac{(\bm{l}_1^2\!-\!\bm{q}^2)\!\langle\lambda'_2|\lambda_2\rangle\!\langle\lambda'_1|\lambda_1\rangle}{\omega_{+,K}\omega_{-,K}\!(\omega_{+,K}\!+\!\omega_{-,K})}\,,\nonumber\\
  V^{\mathrm{Left Triangle}}_{\bar{\p}\p\to\bar{\n}\n}\!\!&=&\!\!-\frac{f_{NN\pi}^2}{4f_0^2}\!\!\int\!\!\frac{\dd^3\bm{l}_1}{(2\pi)^3}\frac{(\bm{l}_1^2\!-\!\bm{q}^2)\!\langle\lambda'_2|\lambda_2\rangle\!\langle\lambda'_1|\lambda_1\rangle}{\omega_{+,\pi}\omega_{-,\pi}\!(\omega_{+,\pi}\!+\!\omega_{-,\pi})}\!-\!\frac{f_{\Lambda NK}^2}{16f_0^2}\!\!\int\!\!\frac{\dd^3\bm{l}_1}{(2\pi)^3}\frac{(\bm{l}_1^2\!-\!\bm{q}^2)\!\langle\lambda'_2|\lambda_2\rangle\!\langle\lambda'_1|\lambda_1\rangle}{\omega_{+,K}\omega_{-,K}\!(\omega_{+,K}\!+\!\omega_{-,K})}\nonumber\\
        \!\!&&\!\!+\frac{f_{\Sigma NK}^2}{16f_0^2}\!\!\int\!\!\frac{\dd^3\bm{l}_1}{(2\pi)^3}\frac{(\bm{l}_1^2\!-\!\bm{q}^2)\!\langle\lambda'_2|\lambda_2\rangle\!\langle\lambda'_1|\lambda_1\rangle}{\omega_{+,K}\omega_{-,K}\!(\omega_{+,K}\!+\!\omega_{-,K})}\,.
\end{eqnarray}
Transforming them into the isospin basis, the potentials are 
\begin{eqnarray}
    V^{I=0}_{\bar{N}N}\!\!&=&\!\!-\frac{3f_{NN\pi}^2}{8f_0^2}\!\!\int\!\!\frac{\dd^3\bm{l}_1}{(2\pi)^3}\frac{(\bm{l}_1^2\!-\!\bm{q}^2)\!\langle\lambda'_2|\lambda_2\rangle\!\langle\lambda'_1|\lambda_1\rangle}{\omega_{+,\pi}\omega_{-,\pi}\!(\omega_{+,\pi}\!+\!\omega_{-,\pi})}\!-\!\frac{3f_{\Lambda NK}^2}{16f_0^2}\!\!\int\!\!\frac{\dd^3\bm{l}_1}{(2\pi)^3}\frac{(\bm{l}_1^2\!-\!\bm{q}^2)\!\langle\lambda'_2|\lambda_2\rangle\!\langle\lambda'_1|\lambda_1\rangle}{\omega_{+,K}\omega_{-,K}\!(\omega_{+,K}\!+\!\omega_{-,K})}\nonumber\\
        \!\!&&\!\!-\frac{3f_{\Sigma NK}^2}{16f_0^2}\!\!\int\!\!\frac{\dd^3\bm{l}_1}{(2\pi)^3}\frac{(\bm{l}_1^2\!-\!\bm{q}^2)\!\langle\lambda'_2|\lambda_2\rangle\!\langle\lambda'_1|\lambda_1\rangle}{\omega_{+,K}\omega_{-,K}\!(\omega_{+,K}\!+\!\omega_{-,K})}\,,\nonumber\\
   V^{I=1}_{\bar{N}N}\!\!&=&\!\!\frac{f_{NN\pi}^2}{8f_0^2}\!\!\int\!\!\frac{\dd^3\bm{l}_1}{(2\pi)^3}\frac{(\bm{l}_1^2\!-\!\bm{q}^2)\!\langle\lambda'_2|\lambda_2\rangle\!\langle\lambda'_1|\lambda_1\rangle}{\omega_{+,\pi}\omega_{-,\pi}\!(\omega_{+,\pi}\!+\!\omega_{-,\pi})}\!-\!\frac{f_{\Lambda NK}^2}{16f_0^2}\!\!\int\!\!\frac{\dd^3\bm{l}_1}{(2\pi)^3}\frac{(\bm{l}_1^2\!-\!\bm{q}^2)\!\langle\lambda'_2|\lambda_2\rangle\!\langle\lambda'_1|\lambda_1\rangle}{\omega_{+,K}\omega_{-,K}\!(\omega_{+,K}\!+\!\omega_{-,K})}\nonumber\\
        \!\!&&\!\!-\frac{5f_{\Sigma NK}^2}{16f_0^2}\!\!\int\!\!\frac{\dd^3\bm{l}_1}{(2\pi)^3}\frac{(\bm{l}_1^2\!-\!\bm{q}^2)\!\langle\lambda'_2|\lambda_2\rangle\!\langle\lambda'_1|\lambda_1\rangle}{\omega_{+,K}\omega_{-,K}\!(\omega_{+,K}\!+\!\omega_{-,K})}\,.
\end{eqnarray}
The isospin factor for the potential due to the left triangle diagram is defined by  
\begin{eqnarray}\label{eq:VTriangle}
    V^{\mathrm{LeftTriangle}}_{\bar{N}N}=-\frac{f^2_{NB_iP}}{32f_0^2}\int\frac{\dd^3\bm{l}_1}{(2\pi)^3}\frac{(\bm{l}_1^2-\bm{q}^2)\!\langle\lambda'_2|\lambda_2\rangle\!\langle\lambda'_1|\lambda_1\rangle}{\omega_{+,P}\omega_{-,P}(\omega_{+,P}+\omega_{-,P})}\mathcal{I}^{\mathrm{LeftTriangle}}_{\bar{N}N\to\bar{N}N}\,,
\end{eqnarray}
where $B_i$ denotes the intermediate baryon. 
The interaction potential from the right triangle diagrams 
are the same as those of the left triangle, in the non-relativistic approximation. 
The isospin factors are listed in Table \ref{tab:Isospinfactor}. 
Comparing with the $NN$ potential given by Ref.~\cite{Haidenbauer:2013oca}, one sees that 
the potential due to $\pi\pi$ exchanges satisfies the G-parity transformation rule, 
as expected, while that for the $K\bar{K}$ exchanges is different.

The box diagrams include planar-box and crossed-box diagrams. For the former, the potentials in 
the physical basis are
 \begin{eqnarray}
     \!\!&&\!\!V^{\mathrm{PlanarBox}}_{\bar{\p}\p\to\bar{\p}\p}=\nonumber\\
     \!\!&&\!\!-\frac{5f_{NN\pi}^4}{8}\!\!\int\!\!\frac{\dd^3 \bm{l}_1}{(2\pi)^3}\frac{(\bm{l}_1^2\!-\!\bm{q}^2)^2\!\langle\lambda'_2|\lambda_2\rangle\!\langle\lambda'_1|\lambda_1\rangle\!-\!4\langle\lambda'_2|(\bm{l}_1\times\bm{q})\cdot\bm{\sigma}|\lambda_2\rangle\!\langle\lambda'_1|(\bm{l}_1\times\bm{q})\cdot\bm{\sigma}|\lambda_1\rangle}{\omega^2_{+,\pi}\omega^2_{-,\pi}(\omega_{+,\pi}+\omega_{-,\pi})}\nonumber\\
    \!\!&&\!\!-\frac{f_{NN\pi}^2f_{NN\eta}^2}{4}\!\!\int\!\!\frac{\dd^3 \bm{l}_1}{(2\pi)^3}\frac{(\bm{l}_1^2\!-\!\bm{q}^2)^2\!\langle\lambda'_2|\lambda_2\rangle\!\langle\lambda'_1|\lambda_1\rangle\!-\!4\langle\lambda'_2|(\bm{l}_1\times\bm{q})\cdot\bm{\sigma}|\lambda_2\rangle\!\langle\lambda'_1|(\bm{l}_1\times\bm{q})\cdot\bm{\sigma}|\lambda_1\rangle}{\omega^2_{+,\eta}\omega^2_{-,\pi}(\omega_{+,\eta}+\omega_{-,\pi})}\nonumber\\
    \!\!&&\!\!-\frac{f_{NN\eta}^4}{8}\!\!\int\!\!\frac{\dd^3 \bm{l}_1}{(2\pi)^3}\frac{(\bm{l}_1^2\!-\!\bm{q}^2)^2\!\langle\lambda'_2|\lambda_2\rangle\!\langle\lambda'_1|\lambda_1\rangle\!-\!4\langle\lambda'_2|(\bm{l}_1\times\bm{q})\cdot\bm{\sigma}|\lambda_2\rangle\!\langle\lambda'_1|(\bm{l}_1\times\bm{q})\cdot\bm{\sigma}|\lambda_1\rangle}{\omega^2_{+,\eta}\omega^2_{-,\eta}(\omega_{+,\eta}+\omega_{-,\eta})}\nonumber\\
    \!\!&&\!\!-\frac{f_{\Lambda NK}^4}{8}\!\!\int\!\!\frac{\dd^3 \bm{l}_1}{(2\pi)^3}\frac{(\bm{l}_1^2\!-\!\bm{q}^2)^2\!\langle\lambda'_2|\lambda_2\rangle\!\langle\lambda'_1|\lambda_1\rangle\!-\!4\langle\lambda'_2|(\bm{l}_1\times\bm{q})\cdot\bm{\sigma}|\lambda_2\rangle\!\langle\lambda'_1|(\bm{l}_1\times\bm{q})\cdot\bm{\sigma}|\lambda_1\rangle}{\omega^2_{+,K}\omega^2_{-,K}(\omega_{+,K}+\omega_{-,K})}\nonumber\\
    \!\!&&\!\!-\frac{5f_{\Sigma NK}^4}{8}\!\!\int\!\!\frac{\dd^3 \bm{l}_1}{(2\pi)^3}\frac{(\bm{l}_1^2\!-\!\bm{q}^2)^2\!\langle\lambda'_2|\lambda_2\rangle\!\langle\lambda'_1|\lambda_1\rangle\!-\!4\langle\lambda'_2|(\bm{l}_1\times\bm{q})\cdot\bm{\sigma}|\lambda_2\rangle\!\langle\lambda'_1|(\bm{l}_1\times\bm{q})\cdot\bm{\sigma}|\lambda_1\rangle}{\omega^2_{+,K}\omega^2_{-,K}(\omega_{+,K}+\omega_{-,K})}\nonumber\\
    \!\!&&\!\!-\frac{f_{\Sigma NK}^2f_{\Lambda NK}^2}{4}\!\!\int\!\!\frac{\dd^3 \bm{l}_1}{(2\pi)^3}\frac{(\bm{l}_1^2\!-\!\bm{q}^2)^2\!\langle\lambda'_2|\lambda_2\rangle\!\langle\lambda'_1|\lambda_1\rangle\!-\!4\langle\lambda'_2|(\bm{l}_1\times\bm{q})\cdot\bm{\sigma}|\lambda_2\rangle\!\langle\lambda'_1|(\bm{l}_1\times\bm{q})\cdot\bm{\sigma}|\lambda_1\rangle}{\omega^2_{+,K}\omega^2_{-,K}(\omega_{+,K}+\omega_{-,K})}\,,\nonumber\\
    \!\!&&\!\!V^{\mathrm{PlanarBox}}_{\bar{\p}\p\to\bar{\n}\n}=\nonumber\\
    \!\!&&\!\!-\frac{f_{NN\pi}^4}{2}\!\!\int\!\!\frac{\dd^3 \bm{l}_1}{(2\pi)^3}\frac{(\bm{l}_1^2\!-\!\bm{q}^2)^2\!\langle\lambda'_2|\lambda_2\rangle\!\langle\lambda'_1|\lambda_1\rangle\!-\!4\langle\lambda'_2|(\bm{l}_1\times\bm{q})\cdot\bm{\sigma}|\lambda_2\rangle\!\langle\lambda'_1|(\bm{l}_1\times\bm{q})\cdot\bm{\sigma}|\lambda_1\rangle}{\omega^2_{+,\pi}\omega^2_{-,\pi}(\omega_{+,\pi}+\omega_{-,\pi})}\nonumber\\
    \!\!&&\!\!-\frac{f_{NN\pi}^2f_{NN\eta}^2}{2}\!\!\int\!\!\frac{\dd^3 \bm{l}_1}{(2\pi)^3}\frac{(\bm{l}_1^2\!-\!\bm{q}^2)^2\!\langle\lambda'_2|\lambda_2\rangle\!\langle\lambda'_1|\lambda_1\rangle\!-\!4\langle\lambda'_2|(\bm{l}_1\times\bm{q})\cdot\bm{\sigma}|\lambda_2\rangle\!\langle\lambda'_1|(\bm{l}_1\times\bm{q})\cdot\bm{\sigma}|\lambda_1\rangle}{\omega^2_{+,\eta}\omega^2_{-,\pi}(\omega_{+,\eta}+\omega_{-,\pi})}\nonumber\\
    \!\!&&\!\!-\frac{f_{\Lambda NK}^4}{8}\!\!\int\!\!\frac{\dd^3 \bm{l}_1}{(2\pi)^3}\frac{(\bm{l}_1^2\!-\!\bm{q}^2)^2\!\langle\lambda'_2|\lambda_2\rangle\!\langle\lambda'_1|\lambda_1\rangle\!-\!4\langle\lambda'_2|(\bm{l}_1\times\bm{q})\cdot\bm{\sigma}|\lambda_2\rangle\!\langle\lambda'_1|(\bm{l}_1\times\bm{q})\cdot\bm{\sigma}|\lambda_1\rangle}{\omega^2_{+,K}\omega^2_{-,K}(\omega_{+,K}+\omega_{-,K})}\nonumber\\
    \!\!&&\!\!-\frac{f_{\Sigma NK}^4}{8}\!\!\int\!\!\frac{\dd^3 \bm{l}_1}{(2\pi)^3}\frac{(\bm{l}_1^2\!-\!\bm{q}^2)^2\!\langle\lambda'_2|\lambda_2\rangle\!\langle\lambda'_1|\lambda_1\rangle\!-\!4\langle\lambda'_2|(\bm{l}_1\times\bm{q})\cdot\bm{\sigma}|\lambda_2\rangle\!\langle\lambda'_1|(\bm{l}_1\times\bm{q})\cdot\bm{\sigma}|\lambda_1\rangle}{\omega^2_{+,K}\omega^2_{-,K}(\omega_{+,K}+\omega_{-,K})}\nonumber\\
    \!\!&&\!\!+\frac{f_{\Lambda NK}^2f_{\Sigma NK}^2}{4}\!\!\int\!\!\frac{\dd^3 \bm{l}_1}{(2\pi)^3}\frac{(\bm{l}_1^2\!-\!\bm{q}^2)^2\!\langle\lambda'_2|\lambda_2\rangle\!\langle\lambda'_1|\lambda_1\rangle\!-\!4\langle\lambda'_2|(\bm{l}_1\times\bm{q})\cdot\bm{\sigma}|\lambda_2\rangle\!\langle\lambda'_1|(\bm{l}_1\times\bm{q})\cdot\bm{\sigma}|\lambda_1\rangle}{\omega^2_{+,K}\omega^2_{-,K}(\omega_{+,K}+\omega_{-,K})}\,.\nonumber\\
 \end{eqnarray}
Transforming them into the isospin basis, one has 
\begin{eqnarray}
    \!\!&&\!\!V^{I=0}_{\bar{N}N}=\nonumber\\
    \!\!&&\!\!-\frac{9f_{NN\pi}^4}{8}\!\!\int\!\!\frac{\dd^3 \bm{l}_1}{(2\pi)^3}\frac{(\bm{l}_1^2\!-\!\bm{q}^2)^2\!\langle\lambda'_2|\lambda_2\rangle\!\langle\lambda'_1|\lambda_1\rangle\!-\!4\langle\lambda'_2|(\bm{l}_1\times\bm{q})\cdot\bm{\sigma}|\lambda_2\rangle\!\langle\lambda'_1|(\bm{l}_1\times\bm{q})\cdot\bm{\sigma}|\lambda_1\rangle}{\omega^2_{+,\pi}\omega^2_{-,\pi}(\omega_{+,\pi}+\omega_{-,\pi})}\nonumber\\
    \!\!&&\!\!-\frac{3f_{NN\pi}^2f_{NN\eta}^2}{4}\!\!\int\!\!\frac{\dd^3 \bm{l}_1}{(2\pi)^3}\frac{(\bm{l}_1^2\!-\!\bm{q}^2)^2\!\langle\lambda'_2|\lambda_2\rangle\!\langle\lambda'_1|\lambda_1\rangle\!-\!4\langle\lambda'_2|(\bm{l}_1\times\bm{q})\cdot\bm{\sigma}|\lambda_2\rangle\!\langle\lambda'_1|(\bm{l}_1\times\bm{q})\cdot\bm{\sigma}|\lambda_1\rangle}{\omega^2_{+,\eta}\omega^2_{-,\pi}(\omega_{+,\eta}+\omega_{-,\pi})}\nonumber\\
    \!\!&&\!\!-\frac{f_{NN\eta}^4}{8}\!\!\int\!\!\frac{\dd^3 \bm{l}_1}{(2\pi)^3}\frac{(\bm{l}_1^2\!-\!\bm{q}^2)^2\!\langle\lambda'_2|\lambda_2\rangle\!\langle\lambda'_1|\lambda_1\rangle\!-\!4\langle\lambda'_2|(\bm{l}_1\times\bm{q})\cdot\bm{\sigma}|\lambda_2\rangle\!\langle\lambda'_1|(\bm{l}_1\times\bm{q})\cdot\bm{\sigma}|\lambda_1\rangle}{\omega^2_{+,\eta}\omega^2_{-,\eta}(\omega_{+,\eta}+\omega_{-,\eta})}\nonumber\\
    \!\!&&\!\!-\frac{f_{\Lambda NK}^4}{4}\!\!\int\!\!\frac{\dd^3 \bm{l}_1}{(2\pi)^3}\frac{(\bm{l}_1^2\!-\!\bm{q}^2)^2\!\langle\lambda'_2|\lambda_2\rangle\!\langle\lambda'_1|\lambda_1\rangle\!-\!4\langle\lambda'_2|(\bm{l}_1\times\bm{q})\cdot\bm{\sigma}|\lambda_2\rangle\!\langle\lambda'_1|(\bm{l}_1\times\bm{q})\cdot\bm{\sigma}|\lambda_1\rangle}{\omega^2_{+,K}\omega^2_{-,K}(\omega_{+,K}+\omega_{-,K})}\nonumber\\
    \!\!&&\!\!-\frac{3f_{\Sigma NK}^4}{4}\!\!\int\!\!\frac{\dd^3 \bm{l}_1}{(2\pi)^3}\frac{(\bm{l}_1^2\!-\!\bm{q}^2)^2\!\langle\lambda'_2|\lambda_2\rangle\!\langle\lambda'_1|\lambda_1\rangle\!-\!4\langle\lambda'_2|(\bm{l}_1\times\bm{q})\cdot\bm{\sigma}|\lambda_2\rangle\!\langle\lambda'_1|(\bm{l}_1\times\bm{q})\cdot\bm{\sigma}|\lambda_1\rangle}{\omega^2_{+,K}\omega^2_{-,K}(\omega_{+,K}+\omega_{-,K})}\,,\nonumber\\
 \!\!&&\!\!V^{I=1}_{\bar{N}N}=\nonumber\\
 \!\!&&\!\!-\frac{f_{NN\pi}^4}{8}\!\!\int\!\!\frac{\dd^3 \bm{l}_1}{(2\pi)^3}\frac{(\bm{l}_1^2\!-\!\bm{q}^2)^2\!\langle\lambda'_2|\lambda_2\rangle\!\langle\lambda'_1|\lambda_1\rangle\!-\!4\langle\lambda'_2|(\bm{l}_1\times\bm{q})\cdot\bm{\sigma}|\lambda_2\rangle\!\langle\lambda'_1|(\bm{l}_1\times\bm{q})\cdot\bm{\sigma}|\lambda_1\rangle}{\omega^2_{+,\pi}\omega^2_{-,\pi}(\omega_{+,\pi}+\omega_{-,\pi})}\nonumber\\
    \!\!&&\!\!+\frac{f_{NN\pi}^2f_{NN\eta}^2}{4}\!\!\int\!\!\frac{\dd^3 \bm{l}_1}{(2\pi)^3}\frac{(\bm{l}_1^2\!-\!\bm{q}^2)^2\!\langle\lambda'_2|\lambda_2\rangle\!\langle\lambda'_1|\lambda_1\rangle\!-\!4\langle\lambda'_2|(\bm{l}_1\times\bm{q})\cdot\bm{\sigma}|\lambda_2\rangle\!\langle\lambda'_1|(\bm{l}_1\times\bm{q})\cdot\bm{\sigma}|\lambda_1\rangle}{\omega^2_{+,\eta}\omega^2_{-,\pi}(\omega_{+,\eta}+\omega_{-,\pi})}\nonumber\\
    \!\!&&\!\!-\frac{f_{NN\eta}^4}{8}\!\!\int\!\!\frac{\dd^3 \bm{l}_1}{(2\pi)^3}\frac{(\bm{l}_1^2\!-\!\bm{q}^2)^2\!\langle\lambda'_2|\lambda_2\rangle\!\langle\lambda'_1|\lambda_1\rangle\!-\!4\langle\lambda'_2|(\bm{l}_1\times\bm{q})\cdot\bm{\sigma}|\lambda_2\rangle\!\langle\lambda'_1|(\bm{l}_1\times\bm{q})\cdot\bm{\sigma}|\lambda_1\rangle}{\omega^2_{+,\eta}\omega^2_{-,\eta}(\omega_{+,\eta}+\omega_{-,\eta})}\nonumber\\
    \!\!&&\!\!-\frac{f_{\Sigma NK}^4}{2}\!\!\int\!\!\frac{\dd^3 \bm{l}_1}{(2\pi)^3}\frac{(\bm{l}_1^2\!-\!\bm{q}^2)^2\!\langle\lambda'_2|\lambda_2\rangle\!\langle\lambda'_1|\lambda_1\rangle\!-\!4\langle\lambda'_2|(\bm{l}_1\times\bm{q})\cdot\bm{\sigma}|\lambda_2\rangle\!\langle\lambda'_1|(\bm{l}_1\times\bm{q})\cdot\bm{\sigma}|\lambda_1\rangle}{\omega^2_{+,K}\omega^2_{-,K}(\omega_{+,K}+\omega_{-,K})}\nonumber\\
    \!\!&&\!\!-\frac{f_{\Sigma NK}^2f_{\Lambda NK}^2}{2}\!\!\int\!\!\frac{\dd^3 \bm{l}_1}{(2\pi)^3}\frac{(\bm{l}_1^2\!-\!\bm{q}^2)^2\!\langle\lambda'_2|\lambda_2\rangle\!\langle\lambda'_1|\lambda_1\rangle\!-\!4\langle\lambda'_2|(\bm{l}_1\times\bm{q})\cdot\bm{\sigma}|\lambda_2\rangle\!\langle\lambda'_1|(\bm{l}_1\times\bm{q})\cdot\bm{\sigma}|\lambda_1\rangle}{\omega^2_{+,K}\omega^2_{-,K}(\omega_{+,K}+\omega_{-,K})}\,.\nonumber\\
\end{eqnarray}
For the crossed-box diagrams, the potentials are as follows
\begin{eqnarray}
        V^{\mathrm{CrossBox}}_{\bar{\p}\p\to\bar{\p}\p}\!\!\!\!&&=\!-\frac{f_{NN\pi}^4}{8}\int\frac{\dd^3 \bm{l}_1}{(2\pi)^3}\frac{\omega_{+,\pi}^2+\omega_{-,\pi}\omega_{+,\pi}+\omega_{-,\pi}^2}{\omega^3_{-,\pi}\omega^3_{+,\pi}(\omega_{-,\pi}+\omega_{+,\pi})}[(\bm{l}_1^2-\bm{q}^2)^2\langle\lambda'_2|\lambda_2\rangle\langle\lambda'_1| \lambda_1\rangle\nonumber\\
    &&+4\langle\lambda'_2|(\bm{l}_1\times\bm{q})\cdot\bm{\sigma}| \lambda_2\rangle\langle\lambda'_1|(\bm{l}_1\times\bm{q})\cdot\bm{\sigma}| \lambda_1\rangle]\nonumber\\
    &&-\frac{f_{NN\pi}^2f_{NN\eta}^2}{4}\int\frac{\dd^3 \bm{l}_1}{(2\pi)^3}\frac{\omega_{+,\eta}^2+\omega_{-,\pi}\omega_{+,\eta}+\omega_{-,\pi}^2}{\omega^3_{-,\pi}\omega^3_{+,\eta}(\omega_{-,\pi}+\omega_{+,\eta})}[(\bm{l}_1^2-\bm{q}^2)^2\langle\lambda'_2|\lambda_2\rangle\langle\lambda'_1| \lambda_1\rangle\nonumber\\
    &&+4\langle\lambda'_2|(\bm{l}_1\times\bm{q})\cdot\bm{\sigma}| \lambda_2\rangle\langle\lambda'_1|(\bm{l}_1\times\bm{q})\cdot\bm{\sigma}| \lambda_1\rangle]\nonumber\\
    &&-\frac{f_{NN\eta}^4}{8}\int\frac{\dd^3 \bm{l}_1}{(2\pi)^3}\frac{\omega_{+,\eta}^2+\omega_{-,\eta}\omega_{+,\eta}+\omega_{-,\eta}^2}{\omega^3_{-,\eta}\omega^3_{+,\eta}(\omega_{-,\eta}+\omega_{+,\eta})}[(\bm{l}_1^2-\bm{q}^2)^2\langle\lambda'_2|\lambda_2\rangle\langle\lambda'_1| \lambda_1\rangle\nonumber\\
    &&+4\langle\lambda'_2|(\bm{l}_1\times\bm{q})\cdot\bm{\sigma}| \lambda_2\rangle\langle\lambda'_1|(\bm{l}_1\times\bm{q})\cdot\bm{\sigma}| \lambda_1\rangle]\,,\nonumber\\
    V^{\mathrm{CrossBox}}_{\bar{\p}\p\to\bar{\n}\n}\!\!\!\!&&=\!\frac{f_{NN\pi}^4}{2}\int\frac{\dd^3 \bm{l}_1}{(2\pi)^3}\frac{\omega_{+,\pi}^2+\omega_{-,\pi}\omega_{+,\pi}+\omega_{-,\pi}^2}{\omega^3_{-,\pi}\omega^3_{+,\pi}(\omega_{-,\pi}+\omega_{+,\pi})}[(\bm{l}_1^2-\bm{q}^2)^2\langle\lambda'_2|\lambda_2\rangle\langle\lambda'_1| \lambda_1\rangle\nonumber\\
    &&+4\langle\lambda'_2|(\bm{l}_1\times\bm{q})\cdot\bm{\sigma}| \lambda_2\rangle\langle\lambda'_1|(\bm{l}_1\times\bm{q})\cdot\bm{\sigma}| \lambda_1\rangle]\nonumber\\
    &&-\frac{f_{NN\pi}^2f_{NN\eta}^2}{2}\int\frac{\dd^3 \bm{l}_1}{(2\pi)^3}\frac{\omega_{+,\eta}^2+\omega_{-,\pi}\omega_{+,\eta}+\omega_{-,\pi}^2}{\omega^3_{-,\pi}\omega^3_{+,\eta}(\omega_{-,\pi}+\omega_{+,\eta})}[(\bm{l}_1^2-\bm{q}^2)^2\langle\lambda'_2|\lambda_2\rangle\langle\lambda'_1| \lambda_1\rangle\nonumber\\
    &&+4\langle\lambda'_2|(\bm{l}_1\times\bm{q})\cdot\bm{\sigma}| \lambda_2\rangle\langle\lambda'_1|(\bm{l}_1\times\bm{q})\cdot\bm{\sigma}| \lambda_1\rangle]\,.
\end{eqnarray}
Transforming them into the isospin basis, one has 
\begin{eqnarray}
    V^{I=0}_{\bar{N}N}\!\!\!&&\!=\frac{3f_{NN\pi}^4}{8}\int\frac{\dd^3 \bm{l}_1}{(2\pi)^3}\frac{\omega_{+,\pi}^2+\omega_{-,\pi}\omega_{+,\pi}+\omega_{-,\pi}^2}{\omega^3_{-,\pi}\omega^3_{+,\pi}(\omega_{-,\pi}+\omega_{+,\pi})}[(\bm{l}_1^2-\bm{q}^2)^2\langle\lambda'_2|\lambda_2\rangle\langle\lambda'_1| \lambda_1\rangle\nonumber\\
    &&+4\langle\lambda'_2|(\bm{l}_1\times\bm{q})\cdot\bm{\sigma}| \lambda_2\rangle\langle\lambda'_1|(\bm{l}_1\times\bm{q})\cdot\bm{\sigma}| \lambda_1\rangle]\nonumber\\
    &&-\frac{3f_{NN\pi}^2f_{NN\eta}^2}{4}\int\frac{\dd^3 \bm{l}_1}{(2\pi)^3}\frac{\omega_{+,\eta}^2+\omega_{-,\pi}\omega_{+,\eta}+\omega_{-,\pi}^2}{\omega^3_{-,\pi}\omega^3_{+,\eta}(\omega_{-,\pi}+\omega_{+,\eta})}[(\bm{l}_1^2-\bm{q}^2)^2\langle\lambda'_2|\lambda_2\rangle\langle\lambda'_1| \lambda_1\rangle\nonumber\\
    &&+4\langle\lambda'_2|(\bm{l}_1\times\bm{q})\cdot\bm{\sigma}| \lambda_2\rangle\langle\lambda'_1|(\bm{l}_1\times\bm{q})\cdot\bm{\sigma}| \lambda_1\rangle]\nonumber\\
    &&-\frac{f_{NN\eta}^4}{8}\int\frac{\dd^3 \bm{l}_1}{(2\pi)^3}\frac{\omega_{+,\eta}^2+\omega_{-,\eta}\omega_{+,\eta}+\omega_{-,\eta}^2}{\omega^3_{-,\eta}\omega^3_{+,\eta}(\omega_{-,\eta}+\omega_{+,\eta})}[(\bm{l}_1^2-\bm{q}^2)^2\langle\lambda'_2|\lambda_2\rangle\langle\lambda'_1| \lambda_1\rangle\nonumber\\
    &&+4\langle\lambda'_2|(\bm{l}_1\times\bm{q})\cdot\bm{\sigma}| \lambda_2\rangle\langle\lambda'_1|(\bm{l}_1\times\bm{q})\cdot\bm{\sigma}| \lambda_1\rangle]\,,\nonumber\\
    V^{I=1}_{\bar{N}N}\!\!\!&&\!=-\frac{5f_{NN\pi}^4}{8}\int\frac{\dd^3 \bm{l}_1}{(2\pi)^3}\frac{\omega_{+,\pi}^2+\omega_{-,\pi}\omega_{+,\pi}+\omega_{-,\pi}^2}{\omega^3_{-,\pi}\omega^3_{+,\pi}(\omega_{-,\pi}+\omega_{+,\pi})}[(\bm{l}_1^2-\bm{q}^2)^2\langle\lambda'_2|\lambda_2\rangle\langle\lambda'_1| \lambda_1\rangle\nonumber\\
    &&+4\langle\lambda'_2|(\bm{l}_1\times\bm{q})\cdot\bm{\sigma}| \lambda_2\rangle\langle\lambda'_1|(\bm{l}_1\times\bm{q})\cdot\bm{\sigma}| \lambda_1\rangle]\nonumber\\
    &&+\frac{f_{NN\pi}^2f_{NN\eta}^2}{4}\int\frac{\dd^3 \bm{l}_1}{(2\pi)^3}\frac{\omega_{+,\eta}^2+\omega_{-,\pi}\omega_{+,\eta}+\omega_{-,\pi}^2}{\omega^3_{-,\pi}\omega^3_{+,\eta}(\omega_{-,\pi}+\omega_{+,\eta})}[(\bm{l}_1^2-\bm{q}^2)^2\langle\lambda'_2|\lambda_2\rangle\langle\lambda'_1| \lambda_1\rangle\nonumber\\
    &&+4\langle\lambda'_2|(\bm{l}_1\times\bm{q})\cdot\bm{\sigma}| \lambda_2\rangle\langle\lambda'_1|(\bm{l}_1\times\bm{q})\cdot\bm{\sigma}| \lambda_1\rangle]\nonumber\\
    &&-\frac{f_{NN\eta}^4}{8}\int\frac{\dd^3 \bm{l}_1}{(2\pi)^3}\frac{\omega_{+,\eta}^2+\omega_{-,\eta}\omega_{+,\eta}+\omega_{-,\eta}^2}{\omega^3_{-,\eta}\omega^3_{+,\eta}(\omega_{-,\eta}+\omega_{+,\eta})}[(\bm{l}_1^2-\bm{q}^2)^2\langle\lambda'_2|\lambda_2\rangle\langle\lambda'_1| \lambda_1\rangle\nonumber\\
    &&+4\langle\lambda'_2|(\bm{l}_1\times\bm{q})\cdot\bm{\sigma}| \lambda_2\rangle\langle\lambda'_1|(\bm{l}_1\times\bm{q})\cdot\bm{\sigma}| \lambda_1\rangle]\,.
\end{eqnarray}
The isospin factors for planar-box and crossed-box diagrams are defined by  
\begin{eqnarray}
    V^{\mathrm{PlanarBox}}_{\bar{N}N}\!\!&=&\!\!-\mathcal{I}^{\mathrm{PlanarBox}}_{\bar{N}N\to \bar{N}N}\frac{f^2_{NB_{i_1} P_1}f^2_{NB_{i_2} P_2}}{8}\times\nonumber\\
     \!\!&&\!\!\int\!\!\frac{\dd^3 \bm{l}_1}{(2\pi)^3}\frac{(\bm{l}_1^2\!-\!\bm{q}^2)^2\langle\lambda'_2|\lambda_2\rangle\!\langle\lambda'_1| \lambda_1\rangle\!-\!4\langle\lambda'_2|(\bm{l}_1\times\bm{q})\cdot\bm{\sigma}|\lambda_2\rangle\!\langle\lambda'_1|(\bm{l}_1\times\bm{q})\cdot\bm{\sigma}|\lambda_1\rangle}{\omega^2_{+,P_1}\omega^2_{-,P_2}(\omega_{+,P_1}+\omega_{-,P_2})}\,,\nonumber\\
    V^{\mathrm{CrossBox}}_{\bar{N}N}\!\!&=&\!\!-\mathcal{I}^{\mathrm{CrossBox}}_{\bar{N}N\to \bar{N}N}\frac{f^2_{NB_{i_1} P_1}f^2_{NB_{i_2} P_2}}{8}\!\!\int\!\!\frac{\dd^3 \bm{l}_1}{(2\pi)^3}\frac{\omega_{+,P_1}^2+\omega_{-,P_2}\omega_{+,P_1}+\omega_{-,P_2}^2}{\omega^3_{+,P_1}\omega^3_{-,P_2}(\omega_{+,P_1}+\omega_{-,P_2})}\times\nonumber\\
    &&[(\bm{l}_1^2-\bm{q}^2)^2\langle\lambda'_2|\lambda_2\rangle\langle\lambda'_1| \lambda_1\rangle+4\langle\lambda'_2|(\bm{l}_1\times\bm{q})\cdot\bm{\sigma}| \lambda_2\rangle\langle\lambda'_1|(\bm{l}_1\times\bm{q})\cdot\bm{\sigma}| \lambda_1\rangle]\,. 
\end{eqnarray}
The isospin factors are listed in Table~\ref{tab:Isospinfactor}. 
In the next section, one can find the final expressions for these potentials,
after the integration has been performed.
Notice that the reducible parts of the box diagrams are not included to avoid double counting once the potential is somehow resummed by the LS equation. See the first two graphs in the third row of Fig.~\ref{fig:NNbarFeynman}. 

 \section{The integration of loop momentum} 
For TBE potentials, the integration of loop momentum in the football, left and right triangle, planar and crossed-box diagrams needs to be dealt with~\cite{Epelbaum2000}.
For the football  diagram, the potential in Eq.~\eqref{eq:VFootball} can be written as
\begin{eqnarray}
    V^{\mathrm{Football}}_{\bar{N}N}=V^{\rm Football}_C\langle\lambda'_2|\lambda_2\rangle\langle\lambda'_1|\lambda_1\rangle\mathcal{I}^{\mathrm{Football}}_{\bar{N}N\to\bar{N}N}\,.
\end{eqnarray}
Here, the regularization of the integration part $V^{\rm Football}_C$ has the following form
\begin{eqnarray}
    V^{\rm Football}_C=\frac{1}{3072\pi^2f_0^4}\left[-4m_P^2-\frac{5}{6}q^2+(6m_P^2+q^2)\left(\frac{R}{2}+\ln\frac{m_P}{\lambda}\right)+w^2(q)L(q)\right]\,,
\end{eqnarray}
where the divergent part is $R=\frac{2}{d-4}+\gamma_E-1-\ln(4\pi)$, and a scale $\lambda$ is introduced in dimensional regularization, 
\begin{eqnarray}
    w(q;m_1,m_2)&=&\frac{1}{q}\sqrt{(q^2+(m_1+m_2)^2)(q^2+(m_1-m_2)^2)}\,,\nonumber\\
    L(q;m_1,m_2)&=&\frac{w(q)}{2q}\ln\frac{(qw(q)+q^2)^2-(m_1^2-m_2^2)^2}{4m_1m_2 q^2}\,.
\end{eqnarray}
For equal mass cases, they can be simplified as
\begin{eqnarray}
   w(q;m,m)&=&\sqrt{q^2+4m^2}\,,\nonumber\\
   L(q;m,m)&=&\frac{w(q)}{q}\ln\frac{w(q)+q}{2m}
\end{eqnarray}
For the triangle diagrams, the potential in Eq.~\eqref{eq:VTriangle} can be written as 
\begin{eqnarray}
    V^{\rm Triangle}_{\bar{N}N}=f_{NB_iP}^2V^{\rm Triangle}_C \langle\lambda'_2|\lambda_2\rangle\langle\lambda'_1|\lambda_1\rangle\mathcal{I}^{\mathrm{Triangle}}_{\bar{N}N\to\bar{N}N}\,,
\end{eqnarray}
where one has, 
\begin{eqnarray}
    V^{\rm Triangle}_C&=&-\frac{1}{768\pi^2f_0^2}\left[-4m_P^2-\frac{13}{3}q^2+(16m_P^2+10q^2)L(q)\right.\nonumber\\
    &&\left.+(18m_P^2+5q^2)\left(R+2\ln\frac{m_P}{\lambda}\right)\right]\,.
\end{eqnarray}
For the planar box diagrams, the potential can be written as
\begin{eqnarray}
    V_{\bar{N}N}^{\rm PlanarBox}&=&f_{NB_{i_1}P_1}^2f_{NB_{i_2}P_2}^2\left(V^{\rm PlanarBox}_{C}\langle\lambda'_2|\lambda_2\rangle\langle\lambda'_1|\lambda_1\rangle\right.\nonumber\\
    &&\left.+V^{\rm PlanarBox}_{S}\langle\lambda'_2|\sigma^i_2|\lambda_2\rangle\langle\lambda'_1|\sigma^i_1|\lambda_1\rangle\right.\nonumber\\
    &&\left.+V^{\rm PlanarBox}_{T}\langle\lambda'_2|\bm{\sigma}_2\cdot \bm{q}|\lambda_2\rangle\langle\lambda'_1|\bm{\sigma}_1\cdot \bm{q}|\lambda_1\rangle\right)\mathcal{I}^{\mathrm{PlanarBox}}_{\bar{N}N\to\bar{N}N}\,,
\end{eqnarray}
where one has 
\begin{eqnarray}
    &&V^{\rm PlanarBox}_{C}\nonumber\\
    &=&\frac{1}{192\pi^2}\left[\frac{5}{3}q^2+\frac{(m_{P_1}^2-m_{P_2}^2)^2}{q^2}+16(m_{P_1}^2+m_{P_2}^2)\right.\nonumber\\
    &&+(23q^2+45(m_{P_1}^2+m_{P_2}^2))\left(R+2\ln\frac{\sqrt{m_{P_1}m_{P_2}}}{\lambda}\right)\nonumber\\
    &&+\frac{m_{P_1}^2-m_{P_2}^2}{q^4}(12q^4+(m_{P_1}^2-m_{P_2}^2)^2-9q^2(m_{P_1}^2+m_{P_2}^2))\ln\frac{m_{P_1}}{m_{P_2}}\nonumber\\
    &&+\frac{2}{w^2(q)}\left(23q^4-\frac{(m_{P_1}^2-m_{P_2}^2)^4}{q^4}+56(m_{P_1}^2+m_{P_2}^2)q^2\right.\nonumber\\
    &&\left.\left.+8\frac{m_{P_1}^2+m_{P_2}^2}{q^2}(m_{P_1}^2-m_{P_2}^2)^2+2(21m_{P_1}^4+22m_{P_1}^2m_{P_2}^2+21m_{P_2}^4\right)L(q)\right]\,,\nonumber\\
    V^{\rm PlanarBox}_{T}&=&-\frac{1}{8\pi^2}\left[L(q)-\frac{1}{2}-\frac{m_{P_1}^2-m_{P_2}^2}{2q^2}\ln\frac{m_{P_1}}{m_{P_2}}+\frac{R}{2}+\ln\frac{\sqrt{m_{P_1}m_{P_2}}}{\lambda}\right]\nonumber\\
    &=&-\frac{1}{q^2}V^{\rm PlanarBox}_{S}\,.
\end{eqnarray}
For the crossed-box diagrams, the potential can be written as
\begin{eqnarray}
    V_{\bar{N}N}^{\rm CrossBox}&=&f_{NB_{i_1}P_1}^2f_{NB_{i_2}P_2}^2\left(V^{\rm CrossBox}_{C}\langle\lambda'_2|\lambda_2\rangle\langle\lambda'_1|\lambda_1\rangle\right.\nonumber\\
    &&\left.+V^{\rm CrossBox}_{S}\langle\lambda'_2|\sigma^i_2|\lambda_2\rangle\langle\lambda'_1|\sigma^i_1|\lambda_1\rangle\right.\nonumber\\
    &&\left.+V^{\rm CrossBox}_{T}\langle\lambda'_2|\bm{\sigma}_2\cdot \bm{q}|\lambda_2\rangle\langle\lambda'_1|\bm{\sigma}_1\cdot \bm{q}|\lambda_1\rangle\right)\mathcal{I}^{\mathrm{CrossBox}}_{\bar{N}N\to\bar{N}N}\,,
\end{eqnarray}
where one has 
\begin{eqnarray}
    V^{\rm CrossBox}_{C}&=&-V^{\rm PlanarBox}_{C}\,,\nonumber\\
     V^{\rm CrossBox}_{T}&=&-\frac{1}{q^2}V^{\rm CrossBox}_{S}=V^{\rm PlanarBox}_{T}\,.
\end{eqnarray}

\bibliographystyle{unsrt}
\bibliography{ref}

\end{document}